\documentclass[review]{elsarticle}
\usepackage{float}
\usepackage{multirow}
\usepackage{amsthm}
\usepackage{graphicx}
\usepackage[export]{adjustbox}
\usepackage{changepage}
\usepackage{url}
\usepackage{amssymb}
\usepackage{ulem}
\usepackage{array}
\usepackage{booktabs}
\usepackage[table]{xcolor}
\usepackage[nointegrals]{wasysym} 
\usepackage{listings}
\usepackage{pdflscape}
\usepackage{rotating}
\usepackage[utf8]{inputenc}
\usepackage{tikz,lipsum,lmodern}
\usepackage[most]{tcolorbox}
\usepackage{pgfplots}
\usepackage[toc,page]{appendix}
\usepackage{textcomp}
\usepackage{epstopdf}
\usepackage{csquotes}
\usepackage{dirtytalk}
\usepackage{parskip}
\usepackage{amssymb}
\usepackage{tikz}
\usepackage{comment}
\usepackage{indentfirst}
\usepackage[colorinlistoftodos,prependcaption,textsize=tiny]{todonotes}
\usepackage[utf8]{inputenc}
\usepackage[english]{babel}
\usepackage{geometry}
\pgfplotsset{compat=1.9}
\usepackage{lineno}
\usepackage{caption} 
\usepackage{adjustbox}
\usepackage{graphicx}
\usepackage{multirow}
\usepackage{pdfpages}
\usepackage{rotating}
\usepackage{ragged2e} 
\usepackage{xurl}
\usepackage[T1]{fontenc}
\usepackage{amsmath}
\usepackage{xspace}
\usepackage{hyperref}

\usepackage{longtable,tabu}
\usepackage{afterpage}
\usepackage{verbatim}
\usepackage{subfigure}
\usepackage{subfloat}
\usepackage{longtable}

\hypersetup
{
  colorlinks   = true, 
  urlcolor     = red,  
  linkcolor    = blue, 
  citecolor    = blue  
}
\journal{Journal of Systems and Software}

\begin{document}
\begin{sloppypar}
\begin{frontmatter}

\title{Characterizing Architecture Related Posts and Their Usefulness in Stack Overflow}




\author[mymainaddress]{Musengamana Jean de Dieu}
\ead{mjados@outlook.com}
\address[mymainaddress]{School of Computer Science, Wuhan University, 430072 Wuhan, China}

\author[mymainaddress]{Peng Liang\corref{mycorrespondingauthor}}
\cortext[mycorrespondingauthor]{Corresponding author at: School of Computer Science, Wuhan University, China. Tel.: +86 27 68776137; fax: +86 27 68776027.}
\ead{liangp@whu.edu.cn}

\author[mysecondaryaddress]{Mojtaba Shahin}
\ead{mojtaba.shahin@rmit.edu.au}
\address[mysecondaryaddress]{School of Computing Technologies, RMIT University, 3000 Melbourne, Australia}

\author[mythirdaddress]{Arif Ali Khan}
\ead{arif.khan@oulu.fi}
\address[mythirdaddress]{M3S Empirical Software Engineering Research Unit, University of Oulu, 90014 Oulu, Finland}

\begin{abstract}
\justifying
\textbf{Context}: Stack Overflow (SO) has won the intention from software engineers (e.g., architects) to learn, practice, and utilize development knowledge, such as Architectural Knowledge (AK). But little is known about AK communicated in SO, which is a type of high-level but important knowledge in development.

\noindent\textbf{Objective}: This study aims to investigate the AK in SO posts in terms of their categories and characteristics as well as their usefulness from the point of view of SO users.

\noindent\textbf{Method}: We conducted an exploratory study by qualitatively analyzing a statistically representative sample of 968 Architecture Related Posts (ARPs) from SO.

\noindent\textbf{Results}: The main findings are: (1) architecture related questions can be classified into 9 core categories, in which “architecture configuration” is the most common category, followed by the “architecture decision” category, and (2) architecture related questions that provide clear descriptions together with architectural diagrams increase their likelihood of getting more than one answer, while poorly structured architecture questions tend to only get one answer.

\noindent\textbf{Conclusions}: Our findings suggest that future research can focus on enabling automated approaches and tools that could facilitate the search and (re)use of AK in SO. SO users can refer to our proposed guidelines to compose architecture related questions with the likelihood of getting more responses in SO. 
\end{abstract}


\end{frontmatter}

 \section{Introduction} \label{introduction} 
Technical Questions and Answers (Q\&A) sites, such as Stack Overflow (SO), have revolutionized how users seek knowledge on the Internet \cite{sadowski2015developers}. SO has shown to be the most prominent community Q\&A site for knowledge sharing and learning in software development, and SO leverages the knowledge and skills of its users, such as developers, to share their thoughts and experience by asking various types of technical questions related to development and providing answers to these questions. Also, SO users can learn novel techniques and tools from SO~\cite{zagalsky2018r}. SO is predominately being used to solve coding problems~\cite{treude2011programmers}, and these problems are often not relevant or less interesting to architects because they focus on lower-level implementation details~\cite{treude2011programmers}. However, ever since this site started growing and being popular, architects have begun to share their competencies, experience, and design problems by asking architecture related questions or providing architecture solutions, such as architecture tactics. In our recent industrial survey on how developers search for architectural information~\cite{musenga2022}, practitioners acknowledged Q\&A sites (e.g., SO) as the most useful source of architectural information (e.g., benefits and drawbacks of architecture solutions). Hence, similar to searching and (re)using existing coding related answers provided in SO to solve programming related problems, software engineers (e.g., architects and developers) also search and (re)use existing architecture solutions in SO for addressing their design concerns. Thus, SO not only accumulates code examples, but also curates a large number of architecture solutions provided to a wide range of architecture related questions or design problems \cite{bi2021mat} \cite{soliman2016architectural}.

Although SO users discuss high-level knowledge in SO, for instance, architecture tactics and quality attributes knowledge \cite{bi2021mat}, architecture knowledge for technology decisions \cite{soliman2016architectural}, to date the majority of the existing studies mainly focus on analyzing programming related knowledge in SO posts from different perspectives. For example, Diamantopoulos and Symeonidis \cite{diamantopoulos2015employing} employed source code information to improve question-answering in SO, and Zhang \textit{et al}. \cite{zhang2018code} investigated the quality of code examples in SO programming related posts. Little work has focused on analyzing architectural knowledge provided in Architecture Related Posts (ARPs) in SO. For instance, Bi \textit{et al}. \cite{bi2021mat} mined posts from SO and structured the design relationships between architectural tactics and quality attributes used in practice. Liu \textit{et al.} \cite{liu2020mining} extracted SO posts and mined the design pattern use scenarios and related design pattern pairs. Soliman \textit{et al.} \cite{soliman2018improving} developed a search approach (i.e., a domain specific search approach) for searching architecture knowledge in SO. In another work, Soliman \textit{et al}. \cite{soliman2021exploring} conducted an empirical study with 50 software engineers, who used Google to make design decisions using Attribute Driven Design \cite{cervantes2016designing}, and they determined how effective web search engines are to find relevant architectural information from various sources (including SO) and to capture AK. Malavolta \textit{et al}. \cite{malavolta2021mining} extracted data from five open source software repositories (including SO), and mined architectural tactics for energy-efficiency applied by practitioners in real robotics projects. Tian \textit{et al}. \cite{tian2019developers} studied SO users’ conception of architectural smells using SO posts. The abovementioned studies extracted ARPs from SO and investigated architecture knowledge (high-level concepts) from different aspects. However, prior work is only based on architecture related questions and their associated answers. In contrast, our work covers the entire ARP, including its question, all comments under the question, all answers associated with the question, and all comments under the answers. In addition, no prior study has specifically investigated architectural knowledge provided in ARPs (answers) with regard to their usefulness. Moreover, there has been no comprehensive research on exploring architectural knowledge communicated by SO users in terms of their types, design contexts, characteristics, and usefulness, which is the focus of this study. Analyzing and understanding how SO users deal with architecture design concerns in online developer communities, such as SO, brings three benefits: (1) it provides key insights about the types of design problems SO users face during their architecture design and the types of architecture solutions discussed as well as their usefulness, (2) it can help to know the design contexts in which architecture problems are raised, and (3) it can help to know the characteristics of architecture problems and solutions discussed. These benefits provide an opportunity to develop new techniques and tools that can help SO users search and (re)use architectural knowledge shared in online developer communities. Therefore, this study aims to complement prior works by analyzing the characteristics and categories of ARPs in SO as well as their usefulness from the point of view of SO users. In this study, we treated usefulness (one quality criterion of posts, e.g., answers, in Q\&A sites \cite{zhu2009multi}) using the definition in \cite{zhu2009multi} (i.e., are the answers useful to address the questions?).

To achieve the \textbf{goal} of this study (see Section \ref{goalandRQs}), we conducted an exploratory study to investigate various aspects (e.g., categories and characteristics) of ARPs in SO. More specifically, we extracted 32,182 posts from SO. We went on to manually filter out irrelevant posts and got 10,423 candidate ARPs. Since 10,423 candidate ARPs were a quite large dataset, and it was not easy to manually analyze this size of dataset with human effort and get accurate and comprehensive results, we used the power statistics and calculated a representative sample size \cite{israel1992determining} of these 10,423 ARPs. With a 95\% confidence level and 3\% margin of error, the final representative sample size calculated was 968 ARPs. Then, we randomly selected 968 ARPs from the 10,423 ARPs and analyzed them for answering a set of research questions (see Table \ref{researchQuestions}). Specifically, we manually analyzed the 968 ARPs using open coding and constant comparison from Grounded Theory (GT) \cite{stol2016grounded} to answer those research questions. The \textbf{main results and findings} of this study are that: (1) SO users ask a broad spectrum of architecture related questions ranging from \textit{architecture tool} to \textit{architecture configuration}, \textit{architecture implementation} to \textit{architecture deployment}. (2) The useful architecture solutions are classified into seven categories as a taxonomy (see Figure \ref{TaxonomyOfArchSolution}), such as \textit{solution for architecture configuration}, \textit{solution for architecture implementation}, \textit{architecture tactic}, and \textit{architecture pattern}. One observation is that the identified categories of these posts (questions and answers) cover almost all the architecting activities that span from the initial stages (i.e., architectural analysis and synthesis \cite{HofmeisterGenModel2007}) of architectural creation as well as the later stages (i.e., architectural implementation and maintenance \& evolution \cite{tang2010comparative}) in a system lifecycle. Thus our identified categories of ARPs can support the mentioned architecting activities during the architecture lifecycle, and SO can be considered as one of the sources of architectural knowledge \cite{soliman2016architectural}. We found that architecture related questions that provide \textit{clear descriptions together with architectural diagrams} increase their likelihood of getting more than one answer, while poorly structured architecture questions tend to only get one answer. (3) SO users frequently use two terms related to usefulness (i.e., \textit{useful} and \textit{helpful}) to explicitly communicate about the usefulness of certain architecture solutions provided to their associated architecture related questions.

This study makes the following three contributions: (1) a classification and characterization of architecture related questions that SO users (e.g., developers) asked in SO; (2) a list of identified design contexts in which architecture related questions were raised; and (3) a classification (i.e., a proposed taxonomy) and characterization of useful architecture solutions in SO. Our study findings can be beneficial to various stakeholders. For example, researchers can refer to our proposed taxonomy of useful architecture solutions in SO as a guidance to develop new automated approaches and tools that could mine and locate architecture solutions (e.g., \textit{solution for architecture configuration}, see Figure \ref{TaxonomyOfArchSolution}) for addressing similar design concerns (e.g., questions that ask about architecture configuration, see Table \ref{CategoriesOfQuestions}). This can facilitate SO users to check the questions and solutions that are relevant to their design concerns. SO can use our results to better adjust its answers and comments organization mechanisms and enhance the search and (re)use of useful architecture solutions in SO.

The rest of this paper is structured as follows: Section \ref{Background} presents the background of the study. Section \ref{researchdesign} describes the research methodology, and Section \ref{results} elaborates the study results. Section \ref{discussionAndImplications} analyzes the results and discusses their implications. Section \ref{ThreatValidity} presents the threats to the validity of the study results, and Section \ref{relatedwork} summarizes the related work. Finally, Section \ref{conclusionFurtureWork} concludes this work with potential areas of future research.

\section{Background}\label{Background}
In this section, we introduce the background concepts used in this study, including Stack Overflow, architecture knowledge, architecture problem, design context, and architecture solution.

\subsection{Stack Overflow} 
Stack Overflow is one of the websites that make Stack Exchange\footnote{\url{https://stackexchange.com/sites}} network, which provides a Q\&A platform for its users to share knowledge across various domains (e.g., programming, design, statistics, mathematics). SO users exchange knowledge related to software development by asking questions or providing answers to existing questions. Among other development knowledge, architecture knowledge, such as drawbacks and benefits of architecture solutions (e.g., patterns and tactics) in certain application domains, has been shared at SO to support architecting activities \cite{hofmeister2007general}\cite{li2013application}. Mining architecture knowledge in Q\&A websites, specifically in SO, has been the subject of the architecture research community in recent years, such as architectural knowledge for technology decisions \cite{soliman2016architectural} and architecture tactics and quality attributes \cite{bi2021mat}, in order to support the architecting process.

\subsection{Architecture knowledge}
Software Architecture (SA) is a set of structures comprising software elements, the relationships among them, and the properties of the elements and relationships \cite{SA2012}. Building an architecture of a software system often requires knowledge, especially architecture knowledge \cite{10yearsSAKM}, and skills. Architecture knowledge, such as architecture decisions and their rationale \cite{jansen2005SoftArch}, benefits and drawbacks of architecture solutions \cite{soliman2016architectural}, is one of the most important types of knowledge in software development \cite{SA2012}. Architectural knowledge is often described in various formats, such as textual and graphical representation \cite{Malavolta2013WhatIN} and this knowledge is recorded in various sources, such as books \cite{SA2012}, technical blogs and tutorials \cite{soliman2021exploring}, developer mailing lists (e.g., ArgoUML \cite{bi2021architecture}), Q\&A sites (e.g., SO \cite{bi2021mat}). In this study, we investigated architecture knowledge discussed in SO from various aspects, such as categories and characteristics of ARPs in SO, SO users' discussions on the usefulness of architecture solutions provided in SO.

\subsection{Architecture problem}
Architecture problems (such as ``\textit{any testable architecture or design pattern for an MFC application?}''\footnote{\url{https://tinyurl.com/2z69uzs5}}) arise during development when addressing specific architecture design concerns (e.g., quality attributes) and their trade-offs \cite{SA2012}. There are various problems related to architecture design that are asked in SO. In this study, we investigated the categories of architecture problems/questions, specifically, the categories of architecture related questions asked in SO (see Section \ref{resultsofRQ1}). 

\subsection{Design context}
Design context of a software system comprises the knowledge that an architect needs to have about the environment (e.g., a hardware platform) in which a system is expected to operate \cite{bedjeti2017modeling}. 
Design contexts can be seen as “\textit{conditions that influence design decisions but are not specified explicitly as requirements}” \cite{tang2008towards}. Harper and Zheng suggested that design contexts are forces that influence stakeholders’ concerns \cite{harper2015exploring}. 
There are some works that categorize design contexts. Bedjeti \textit{et al.} identified four context categories of an architecture viewpoint (i.e., platform context, user context, application context, and organizational context) \cite{bedjeti2017modeling}. Petersen and Wohlin provided a checklist for documenting design contexts from six perspectives: product, processes, practices and techniques, people, organization, and market \cite{petersen2009context}. 
Groher and Weinreich studied environmental factors that influence architecture decision making~\cite{groher2015study}, and they identified eight categories: company size, business factors, organizational factors, technical factors, cultural factors, individual factors, project factors, and decision scope. In our study, we investigated the design contexts that were discussed in architectural related posts in SO, and we referred to the classification of design contexts proposed in two existing studies \cite{bedjeti2017modeling}\cite{petersen2009context} (see Section \ref{resultsofRQ2}).

\subsection{Architecture solution}
Architecture solutions are the fundamental building blocks in modern software design and they are used to address architecture design concerns \cite{SA2012}. There are various architecture solutions, such as patterns, tactics, and frameworks for addressing different design concerns. \textit{Architecture patterns} (e.g., Model–View–Controller, Client-Server, Publish-Subscribe patterns) are reusable solutions to commonly occurring problems in architecture design within given contexts \cite{SA2012}. Contrarily to changing implementation (e.g., low-level code), once an architecture solution (e.g., an architecture pattern) is adopted and implemented, it is quite difficult and costly to change it~\cite{SA2012}. Architecture patterns determine the overall structure and behavior of a software system \cite{buschmann1996pattern} and are typically selected early during development for achieving multiple system requirements (e.g., quality attributes) \cite{SA2012}. 
In this study, we studied SO users' discussions on the usefulness of architecture solutions, for example, patterns, tactics, frameworks (see Section \ref{resultsofRQ5}), as well as the categories and characteristics (in Section \ref{resultsofRQ6}) of architecture solutions that were considered useful in SO.

\section {Research design} \label{researchdesign}

We carried out an exploratory study on various aspects (e.g., categories and characteristics) of ARPs in SO. In the following subsections, we describe the details of the research design of this study, including the goal and Research Questions (RQs) in Section \ref{goalandRQs} and the execution of this study in Section \ref{studyExcution}.

\subsection{Goal and research questions} \label{goalandRQs}
 The overall \textbf{goal} of this study based on Goal-Question-Metric approach \cite{caldiera1994goal} is “{\textit{to \textbf{analyze} the ARPs (questions and answers) in SO \textbf{for the purpose of} investigating their categories, characteristics, and usefulness \textbf{from the point of view of} SO users \textbf{in the context of} software development in practice}}”. Following the goal of this research, we derived six research questions (see Table \ref{researchQuestions}) that aim to examine four aspects that highlight the question and answer threads of SO posts, including (1) categorization of architecture related questions, (2) the design contexts in which architecture related questions were raised, (3) characterization of architecture related questions that have more than one answer and characteristics of architecture related questions that only have one answer, and (4) categorization and characterization of architecture solutions that are considered useful.

\begin{table} 
\small
	\caption{Research questions and their rationale} 
	\label{researchQuestions}
	\begin{tabular}{p{5cm}p{10.5cm}}
		\toprule
	\textbf{Research Question}& \textbf{Rationale}\\
		\midrule
	 \textbf{RQ1.} What architecture related questions are asked in SO? & Architecture related questions (design problems) are mainly asked to address certain design concerns (e.g., quality attributes of a system) during architecting activities, for example, architectural analysis. SO curates different types of architecture related questions that are raised with various design issues. The answer to this RQ can help researchers to be aware of the areas of interest of SO users in architecture design and help practitioners to get an insight into the architecture related questions asked in SO so that they can provide practical contributions.\\
  
	 \textbf{RQ2.} What are the design contexts in which architecture related questions were raised? & Design contexts comprise the knowledge about the environments in which systems are expected to operate \cite{bedjeti2017modeling}. Design contexts are indispensable ingredients that can drive the architecture design of a system \cite{bedjeti2017modeling}. A system of similar functionalities can operate differently in different contexts \cite{petersen2009context}. Although the importance of considering design contexts during architecture design has been recognized, there is limited understanding on what design contexts are considered in architecture design. The answer to this RQ can help researchers and practitioners be aware of typical design contexts in which architecture related questions are raised in SO.\\
	 
	 \textbf{RQ3.} What are the characteristics of architecture related questions in SO that have more than one answer? & One major challenge during architecture design is choosing the right architecture solutions to address the requirements of the systems~\cite{soliman2015enriching}. Although different architecture solutions act as alternative solutions to similar architecture problems, they differ in terms of their qualities \cite{soliman2015enriching}. Therefore, providing more than one answer (e.g., alternative solutions) to architecture problems/questions is important as they provide a wide range of possibilities for making architecture design decisions. With this RQ, we identify and examine the characteristics of architecture related questions that get more than one answer. By characteristics of an architecture related question, we mean certain features, such as architectural diagrams, in the content of the question or question formulation \cite{WritingGoodquest2018}, that distinguish the architecture related question to another or make the architecture related question get attraction from SO users and get more than one answer. The answer to this RQ can help researchers and practitioners know what motivates SO users to provide more solutions to these questions, and consequently improve or prevent unanswered architecture related questions in SO.\\
	 
	 \textbf{RQ4.} What are the characteristics of architecture related questions in SO that only have one answer? & Some architecture related questions fail to continuously get attention from SO users by answering them. Similar to RQ3, we want to examine the factors behind this situation. We study the characteristics of architecture related questions that only get one answer. The answer of this RQ can help researchers and practitioners know what demotivates SO users to continue answering these questions and design general guidelines for SO users to compose architecture related questions with the likelihood of getting more responses in SO.\\
	 
	 \textbf{RQ5.} What are the types of architecture solutions provided in SO that are considered useful by SO users? & There are many architecture solutions (e.g., tactics and patterns) to address architecture related questions provided in SO. However, the quality of solutions/answers provided in SO has been a major concern for researchers and practitioners. As elaborated in the related work (see Section \ref{relatedwork}), this is evident in the growing number of studies, in which the focus is on analyzing the quality of the content in SO posts from different perspectives, for example, code and text. The results of this RQ can help researchers and practitioners be aware of types (a taxonomy) of architecture solutions considered useful in SO.\\
	 
	 \textbf{RQ6.} What are the characteristics of architecture solutions in SO that are considered useful by SO users? & Zhang \textit{et al}. \cite{zhang2018code} argued that accepted, highly voted, and frequently viewed SO posts are not always reliable or useful in SO. Identifying the features of architecture solutions that are considered useful helps to better understand what SO users consider when accepting architecture solutions as useful ones, thus providing insights for improving the current answering mechanism of architecture related questions and helping SO users retrieve their desired architecture solutions.\\ 
		\bottomrule 
	\end{tabular}
	\label{table:1}
\end{table}

\subsection{Study Execution} \label{studyExcution}
\begin{figure}[H] 
 \centering
 \includegraphics[scale=0.6]{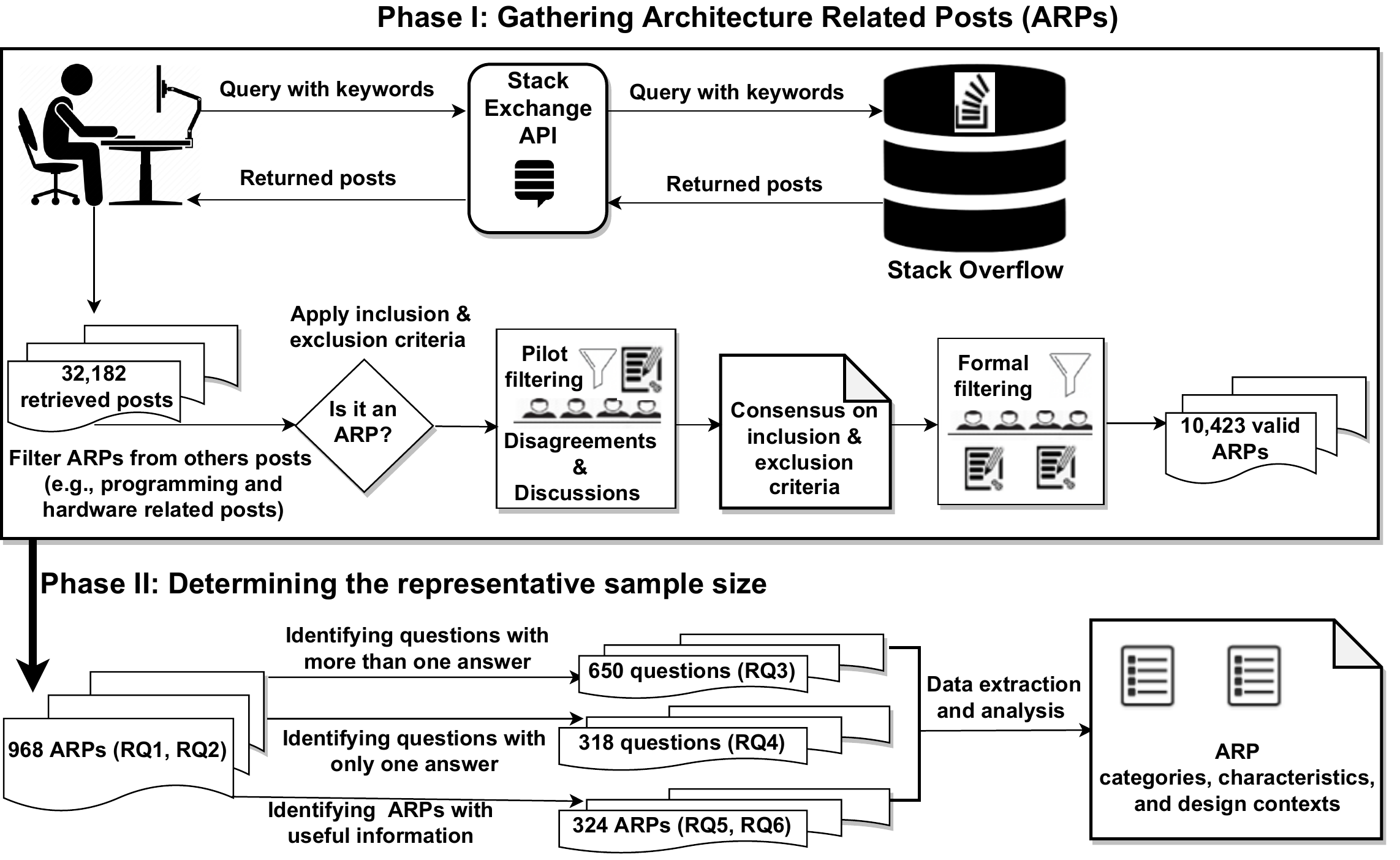}
 	\caption{An overview of data collection and analysis}
 \label{DrawingStudyExcution}
\end{figure}

 In this subsection, we describe the process of data collection and analysis of ARPs. Figure \ref{DrawingStudyExcution} shows an overview of the two processes (i.e., data collection and analysis).

\subsubsection{Data collection} \label{datacollection}
Our data collection is divided into two phases, namely \textit{Phase I: Gathering architecture related posts} and \textit{Phase II: Determining the representative sample size}, as detailed below: 

\textit{Phase I: Gathering architecture related posts} \label{PhaseI}

\textit{a) Search terms}:\label{searchquery} 
Before we decided the most suitable terms for capturing posts relevant to architecture design, we first performed a pilot search with several terms, namely “architect*” (i.e., “architect”, “architecture”, “architectural”, and “architecting”) and “design*” (i.e., “design” and “designing”), within SO. The process was carried out by using a SQL query through the query interface provided by StackExchange Data Explorer\footnote{\url{https://data.stackexchange.com/stackoverflow/query/new}}, which is a web interface that allows the execution of SQL queries on data from Q\&A sites, including SO. After the pilot search with the mentioned terms, we saw that SO users mostly use the terms “design*” (i.e., “design” and “designing”) in the programming context in SO, for instance, “singleton design pattern”\footnote{\url{https://tinyurl.com/8yks7nhm}}. Moreover, we were aware that Soliman \textit{et al}. \cite{soliman2016architectural} identified distinctive terms between ARPs and pure programming posts from their studied sample of SO posts. However, in our study, we did not use those distinctive terms to search ARPs in SO due to the following two reasons: 

(1) The purposes of our work and Soliman \textit{et al}.'s work in \cite{soliman2016architectural} are different. The purpose of the work in \cite{soliman2016architectural} is technology related architecture knowledge extraction from SO. Specifically, the authors in \cite{soliman2016architectural} identified and analyzed ARPs that mainly discuss architectural knowledge for technology decisions, such as the pros and cons of a technology solution in a certain application. In addition, the authors in \cite{soliman2016architectural} claimed that they did not find many pure architectural concepts (such as architectural pattern or tactic) in their dataset of ARPs. In contrast, our study takes the problems from a wider scope. Specifically, our study aims to identify and analyze ARPs from SO by looking at various architectural information, including architecture patterns, tactics. Therefore, using the specific distinctive terms, such as versus, alternative, pros, cons, xmpp, that were found in \cite{soliman2016architectural} may lead to missing other relevant ARPs, which may affect the completeness of the retrieved ARPs. 

(2) Using the distinctive terms found in Soliman \textit{et al}.'s work~\cite{soliman2016architectural} may lead to bias in the search results. The relevancy and completeness of extracted ARPs may affect the correctness of the answers to our six RQs (see Table \ref{researchQuestions}). Thus, including the specific distinctive terms that were found in \cite{soliman2016architectural} in the search queries may lead to the situation that the search results are biased to those terms. 

Therefore, we selected the general terms “architect*” (i.e., “architect”, “architecture”, “architectural”, and “architecting”) to be used in our search. It is worth mentioning that we did not use the search terms to search exclusively through tags only because tags can sometimes be less informative and ineffective \cite{barua2014developers}. There are several disadvantages of using tags as the only approach to determine whether a post is related to a topic. This is due to the reason that a user who created a post could be unsure about the title of the most appropriate tag for their discussion, which can lead to the use of incorrect or irrelevant tags \cite{tahir2018can}. For example, in this architecture related post\footnote{\url{https://tinyurl.com/mb9y37z4}} that asked for an architecture pattern that can be used in the design of a single webform application, a developer used tags (“jc\#”, “asp.net”, and “web”), and these tags cannot immediately tell in which contexts (e.g., architecture or programming context) they are really used. Another problem with user-defined tags is that users may try to add as many tags as possible (SO allows up to 5 tags) to raise the number of views and probably increase the probability of getting responses quickly \cite{allamanis2013and}. Thus, while tags can be helpful to capture posts related to architecture design, using tags exclusively may miss important posts on this topic. Hence, we decided to add the title and body of the questions into the search. For example, by following the criteria of the query interface provided by Stack Exchange\footnote{\url{https://data.stackexchange.com/stackoverflow/query/new}}, for the term “architect”, we searched in the title, tags, and body of posts by using this query: \texttt{SELECT p.Id, p.Tags, p.Title, p.Body as “Questions Body”, p.Score as “Questions Score”, p.Answercount as “Answer Count” FROM Posts p WHERE (p.Body like ‘\%architect\%' or p.Title like ‘\%architect\%' or p.Tags like ‘\%architect\%') AND p.Score \textgreater 0 and p.AnswerCount \textless \textgreater 0 ORDER BY p.Score DESC}. In our replication package~\cite{dataset}, we provided the complete SQL query used to search ARPs in SO, such as how the title, tags, and body of a post were combined during the search. The searching process resulted in 32,182 posts (see Figure \ref{DrawingStudyExcution}). Note that we used the mentioned search terms (i.e., “architect”, “architecture”, “architectural”, and “architecting”), not for the purpose of accumulating all ARPs in SO, but for gathering sufficient data for a relatively comprehensive analysis to achieve the goal of this study. 

\textit{b) Filtering ARPs from other posts (i.e., programming and hardware related posts)}:\label{Filtering ARPs}
We found that SO users use the term “architecture” not only in the context of software architecture, but also in other contexts, such as hardware architecture context (e.g., ARM6 CPU architecture\footnote{\url{https://tinyurl.com/f8sjvzz2}}) and programming context (e.g., array architecture\footnote{\url{https://tinyurl.com/3ps7b3ek}}), when describing their concerns in the SO posts. Therefore, we need to filter the retrieved 32,182 posts and exclude those posts related to programming and hardware architecture. To do so, we performed context analysis and applied our defined inclusion and exclusion criteria (see Table \ref{mainInclusionExclusion}) to accurately filter and separate software ARPs from other types of posts mentioned above.

Before the formal post filtering (manual inspection), to reach an agreement about the inclusion and exclusion criteria (see Table \ref{mainInclusionExclusion}), a pilot filtering was performed whereby the first author took a random sample of 1,000 posts from the 32,182 posts. He manually checked them with our defined criteria (see Table \ref{mainInclusionExclusion}). The other three authors checked and examined the results so that all the authors (four authors) of this study could get a consensus on the understanding of the defined inclusion and exclusion criteria. Thereafter, we got controversy and misunderstanding on 51 posts from the filtered results. Such controversy and misunderstanding were discussed between the four authors of this study till a consensus was reached. The first author carried on with the formal post filtering based on the inclusion and exclusion criteria. The process continued till all the 32,182 posts were manually checked. This step resulted in 10,423 candidate ARPs (see Figure \ref{DrawingStudyExcution}). The results from this round were checked and verified by the other three authors of this study, and it took us twenty one full days to identify and separate ARPs from programming and hardware architecture related posts in these 32,182 posts.

\begin{table} [h!] 
\small
\captionsetup{font=scriptsize}
	\caption{Inclusion and exclusion criteria for filtering ARPs from programming and hardware related posts}
	\label{mainInclusionExclusion}
	\begin{tabular}{p{15cm}}
		\toprule
	\textbf{Inclusion criteria}\\
	\midrule
	\textbf{I1.} An ARP should contain a discussion on software architecture, for example, architecture design and architecture tactics.\\
	 
	\textbf{I2.} An ARP should contain at least one answer attached to its architecture related question as we aim to study the factors that make these questions have more than one answer or only have one answer and the usefulness of their answers.\\
	 
	\textbf{I3.} An ARP should contain at least one data item that can be extracted according to the data items defined in Table \ref{dataExtraction}.\\
	
	\midrule
	\textbf {Exclusion criteria}
	\\ \midrule
    \textbf{E1.} An ARP that has a score (i.e., medium number of down/upvote) that is less than 1 is excluded since we want to make sure that all studied posts have attracted enough attention from the community \cite{UnderstQuestQuali2014}.\\
    \bottomrule 
	\end{tabular}
\end{table}

\textit{Phase II: Determining the representative sample size}\label{PhaseII_sampleSize}

The 10,423 candidate ARPs (filtered from the previous phase (i.e., Phase I)) are a quite large dataset, and it is not realistic to analyze this size of dataset with human effort and get accurate and comprehensive results. Thus, in order to get statistically significant results, we used the power statistics and calculated a representative sample size \cite{israel1992determining} of these 10,423 ARPs. At a confidence level of 95\%, we set a margin of error (i.e., how much we can expect our analysis results to reflect the view of the overall dataset) to 3\% for the whole 10,423 ARPs. The final representative sample size calculated is 968 ARPs. Then, we randomly selected 968 ARPs from the 10,423 ARPs and analyzed them for answering the six RQs (see Table \ref{researchQuestions}). To be more specific, except for RQ1 and RQ2 on which we used 968 (a representative sample size of ARPs) to answer them, we used subsets of the 968 ARPs that satisfy our defined criteria (in the following steps) with respect to the purposes of the remaining RQs (i.e., RQ3, RQ4, RQ5, and RQ6) (see Table \ref{researchQuestions}). We followed the following steps to further divide our calculated representative sample size of ARPs (968) into subsets of the ARPs that are relevant to answering the remaining RQs:

\textit{\textbf{Step 1}: Identification of questions that have more than one answer (RQ3) and questions that only have one answer (RQ4)}. 
As stated in the rationale of these two RQs (see Table \ref{researchQuestions}), we want to examine the factors that make such architecture questions get more than one answer or only get one answer. Thus we considered comments posted on architecture related questions by referring to these two studies \cite{obsolete2019}\cite{readingAnswers2019}, in which the authors argued that the quality of an answer is a combination of both the answer and its associated comments as comments may provide additional information about the answer, for example, improvement of answers \cite{readingAnswers2019} and obsoleted answers \cite{obsolete2019}. Therefore, in our study, we included comments posted on architecture related questions and studied what makes these posts get more than one answer (RQ3) or only get one answer (RQ4). More specifically, the first author manually checked 968 ARPs and their comments, and got 650 architecture related questions (wherein each question has more than one answer). These questions were used to answer RQ3 (see Figure \ref{DrawingStudyExcution}). On the other hand, the first author followed the same procedure (manual inspection of the 968 ARPs for RQ3) and got 318 architecture related questions (wherein each question has only one answer). These questions were used to answer RQ4 (see Figure \ref{DrawingStudyExcution}).

\begin{figure}[h!]
 \centering
 \fbox{\includegraphics[scale=0.45]{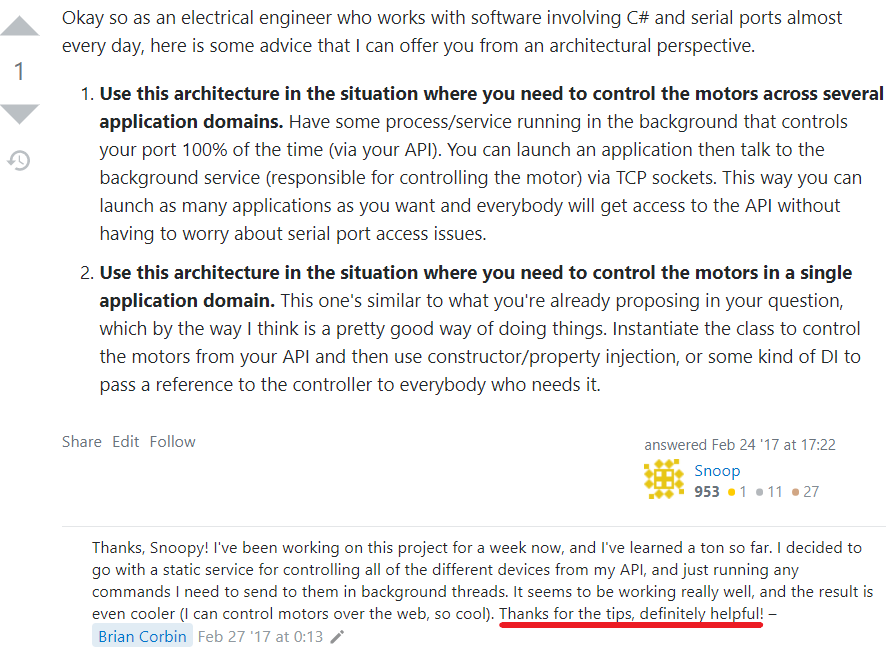}} 
 	\caption{An example answer that was commented to be helpful}
	\label{UsefulComment}
\end{figure}

\textit{\textbf{Step 2}: Identification of ARPs with useful knowledge}. For answering RQ5 and RQ6, we need to identify ARPs with useful knowledge from the representative random sample of ARPs (968 ARPs).
We referred to these two studies (i.e., \cite{obsolete2019, readingAnswers2019}) to identify ARPs with usefulness knowledge. These studies by Zhang \textit{et al}. \cite{obsolete2019, readingAnswers2019} argued that the quality of an answer is a combination of both the answer and its associated comments as comments may provide additional information to support the answer, such as improvement of answers \cite{readingAnswers2019} and obsoleted answers \cite{obsolete2019}. Therefore, in this study, we included the information in comments to gain a deep understanding of how SO users discuss the usefulness of architecture solutions provided to their architecture related questions in SO. In this study, we did not consider vote score for answering RQ5 and RQ6 since even highly voted SO posts are not always reliable or useful as argued by Zhang \textit{et al.} \cite{zhang2018code}. We observed that SO users occasionally use terms related to usefulness, such as “helpful”, in the comments to indicate that certain architecture solutions provided to their architecture related questions are useful (see Figure \ref{UsefulComment}). Thus, based on this observation along with the aid of our defined selection criteria in Table \ref{subInclusionExclusion}, we manually checked and filtered the 968 ARPs to identify solution threads with useful knowledge (each solution thread includes all solutions to a question (i.e., accepted \& not-accepted solutions) and all the comments that are associated with the solutions. Specifically, to filter out ARPs that do not discuss the usefulness of architecture solutions and reach an agreement about the criteria defined in Table \ref{subInclusionExclusion}, two authors (i.e., the first and second authors) did a pilot ARPs filtering. They independently and manually examined a random sample of 20 ARPs from the 968 ARPs. Similar procedure (i.e., selecting a random sample of data from a large dataset and subsequent manual filtering) has also been employed in recent studies, such as \cite{obie2022violation}. To measure the inter-rater agreement between the first two authors, we calculated the Cohen’s Kappa coefficient \cite{cohen1960coefficient} and got an agreement of 0.898. Note that before a solution was finally included as a relevant one (i.e., a useful solution), the first and second authors first read the solution that was commented to be useful/helpful in order to verify if it is really useful to address the question. Disagreements on the ARPs were discussed between the two authors till a consensus was reached. Then the first author carried on to check and filter the remaining ARPs. The number of resulting ARPs (with useful knowledge) that were used to answer the two RQs (RQ5 and RQ6) is 324 ARPs (see Figure \ref{DrawingStudyExcution}).

\begin{table} [h!] 
\small
	\caption{Inclusion and exclusion criteria for identifying ARPs with useful knowledge}
	\label{subInclusionExclusion}
	\begin{tabular}{p{15cm}}
		\toprule
	\textbf{Inclusion criterion}\\ 
	\midrule
	 \textbf{I1.} A comment in an answer thread must contain one of the keywords related to usefulness, such as “useful”, “helpful”, “beneficial”, “handy”, and “effective”, and this comment is used to signify the usefulness of the answer.\\
	 
	 \midrule
	\textbf {Exclusion criteria}\\ 
	\midrule
	 \textbf{E1.} The keyword related to usefulness, for example, “useful”, “helpful”, “beneficial”, is used to talk about something else (e.g., a question or answer itself is related to a “usefulness” topic rather than being a sign that the answer is likely useful).\\
	 
	 \textbf{E2.} An ARP with controversy discussions on the answer (i.e., if there are two comments in the same answer thread, and one states the usefulness of the answer while another states its uselessness) is not included.\\
	\bottomrule 
	\end{tabular}
	\label{table3}
\end{table}

\subsubsection{Data extraction and analysis} \label{dataExtractionAndAnaly} 

(1) Data extraction: We performed the data extraction process by identifying the relevant information to be extracted from 968 ARPs to answer our defined RQs (see Table \ref{researchQuestions}). In Table \ref{dataExtraction}, we present the data items for which the relevant information was extracted from the candidate ARPs. It also shows the RQs that are supposed to be answered using the extracted data. The data extraction was subsequently followed by data analysis, and these two processes were conducted and recorded with the aid of MAXQDA (a qualitative data analysis tool)\footnote{\url{https://www.maxqda.com/}}. 

\begin{table} [h!]
\small
\caption{Data items to be extracted from the ARPs with their description, analysis approaches, and relevant RQs} \label{dataExtraction}
	\begin{tabular}{p{0.3cm}p{2.8cm}p{4cm}p{4.5cm}p{1.6cm}}
		\toprule
		\#& Data item                   & Description                                                           & Data analysis approach        & RQs\\
		\midrule
		D1 & Content of the question    & The main content of the question in the ARP & Open coding \& constant comparison                       & RQ1\\
		
		D2 & Design context             & The design context elaborated in the content of the question in the ARP & Predefined classifications in \cite{bedjeti2017modeling} and \cite{petersen2009context}                                            & RQ2\\
		
		D3 & Content of the question and its comments & The main content of the question and a summary of the question’s comments in the ARP & Open coding \& constant comparison                                                                                                        & RQ3, RQ4 \\
		
		D4 & Content of the answer     & The main content of the answer in the ARP & Open coding \& constant comparison                           & RQ5\\
		
		D5 & Content of the answer and its comments & The main content of the answer and a summary of the answer's comments in the ARP & Open coding \& constant comparison                                                                                                              & RQ6\\
		\bottomrule 
	\end{tabular}
\end{table}

{(2) Data analysis}: \label{dataAnalyis}

Similarly to several existing studies (e.g., \cite{musenga2022}), we used open coding \& constant comparison to answer RQ1 and RQ3-RQ6 in our study. Open coding \& constant comparison are two widely used techniques from Grounded Theory \cite{stol2016grounded} during qualitative data analysis. Grounded Theory (GT) is a bottom-up approach and focuses on theory generation, rather than extending or verifying existing theories \cite{stol2016grounded}. Open coding generates codes for incidents that can be further classified into concepts and categories \cite{stol2016grounded}. Constant comparison is a continuous process for verifying the generated concepts and categories. Both concepts and categories evolve and saturate until they fit the data \cite{stol2016grounded}. Thus, in this study, we employed open coding \& constant comparison techniques from GT to generate the concepts and categories for answering RQ1 and RQ3-RQ6. Specifically, we used open coding to encode the extracted data items for RQ1 and RQ3-RQ6 (see Table \ref{dataExtraction}) to generate codes. Afterwards, we applied constant comparison to compare the codes identified in one piece of data with the codes that emerge from other data to identify the codes which have similar semantic meanings. We proceeded to group similar codes into high-level concepts and categories. On the other hand, we employed predefined classifications of design contexts in \cite{bedjeti2017modeling} and \cite{petersen2009context} to answer RQ2. We followed the same procedure (encoding and grouping similar codes into high-level categories) to answer RQ2.

Before the formal data analysis, the first author conducted a pilot data analysis for each RQ. Specifically, this analysis process involved the following steps: (1) The first author selected a random set of 100 ARPs from the representative sample size calculated (i.e., 968 ARPs). (2) The first author coded the extracted data (see Table \ref{dataExtraction}) for each RQ. When such posts were unclear and the first author got confused while coding the extracted data, physical meetings with the second author were scheduled to solve such confusion. (3) The first author applied constant comparison and grouped all the codes into higher-level concepts and turned them into categories and subcategories. The grouping process was iterative, in which the first author continuously went back and forth between the concepts, categories, subcategories, and contents of the questions, answers, and comments to revise and refine the concepts, categories, and subcategories. (4) Thereafter, other three authors (the second, third, and fourth authors) checked and validated the results from the pilot data analysis (i.e., concepts, categories, and subcategories). The disagreements were resolved in a meeting using a negotiated agreement approach~\cite{campbell2013coding} to improve the reliability of the pilot data analysis results. The first author carried on with the formal data analysis and followed similar steps used during the pilot data analysis. In the following paragraphs, we provide details of the formal data analysis process: 

\textit{a) For analyzing RQ1, RQ3, and RQ4} 

As abovementioned, we used open coding and constant comparison \cite{stol2016grounded} to manually analyze the extracted data (i.e., content of the question for RQ1 and content of the question and its comments for RQ3 and RQ4) as shown in Table \ref{dataExtraction}. With these RQs, we investigated architecture related questions from two aspects, namely categorization (RQ1) and characterization (RQ3 and RQ4) of these questions (see Table \ref{researchQuestions}). Specifically, regarding the categorization of the questions, the first author studied the content of each architecture related question (from the ARPs that are relevant to answer RQ1 (see Figure \ref{DrawingStudyExcution})) by exploring and identifying their main purposes (e.g., design concerns), such as asking for help on how to refactor the architecture of a system (e.g., refactoring of circular dependencies). Thereafter, the first author summarized each question's purpose in a short sentence. Firstly, the first author went on to encode the summarized sentence. This process was iterative, in which he continuously applied this technique till all ARPs in the dataset were encoded. Secondly, the first author applied constant comparison to compare the codes identified in one summarized sentence with the codes that emerged from other summarized sentences to check the codes which have similar semantic meanings. The first author proceeded to group similar codes into high-level concepts, categories and subcategories. The grouping process was iterative, in which the first author continuously went back and forth between the concepts, categories, subcategories, and contents of the questions to revise and refine both the concepts, categories and their subcategories. To mitigate the personal bias during the formal data analysis, the other authors (second, third, and fourth authors) of this study participated in the validation of the generated codes, concepts, categories, and subcategories. The disagreements were resolved in a meeting using the negotiated agreement approach \cite{campbell2013coding} to improve the reliability of the analysis results for RQ1 as during the pilot data analysis. 
We finally got 9 high-level categories and 21 subcategories as the results of RQ1, and these results are fully elaborated in Section \ref{resultsofRQ1}.

As mentioned in Section \ref{PhaseII_sampleSize}, for answering RQ3 and RQ4, we manually checked the 968 ARPs to identify ARPs with more than one answer (RQ3) and ARPs with only one answer (RQ4). This led to two subsets of the 968 ARPs (i.e., 650 ARPs and 318 ARPs) relevant to answering RQ3 and RQ4 (see Figure \ref{DrawingStudyExcution}). Specifically, in the formal data analysis, the first author analyzed the questions and their attached comments (from the ARPs of the two mentioned subsets, 650 ARPs and 318 ARPs, of the 968 ARPs) by encoding the extracted data for the two RQs (i.e., content of the question and comments as shown in Table \ref{dataExtraction}). He wanted to study if there might be such factors, for example, question formulation \cite{WritingGoodquest2018} or certain features in the question (e.g., architecture diagram) that contribute to such architecture related questions having more than one answer or only having one answer. For example, when investigating RQ3 (questions with more than one answer), one community member posted a comment under an architecture related question saying that “\textit{+1 great question, very well-articulated}”\footnote{\url{https://tinyurl.com/dfw27h38}}, for this comment, he picked a phrase (i.e., a summary of that comment) “well-articulated”. Subsequently, he went on to study the content of the question (e.g., how architectural information is stated in the question) under which this comment was commented, and then he came up with one code that fits this question. He followed the same processes (e.g., grouping similar codes into high-level concepts, categories) that the used when analyzing RQ1 to analyze these two RQs (RQ3 and RQ4). To mitigate the personal bias, the results from this analysis were checked and validated by other three authors of this study. As in the analysis of RQ1, we held a meeting and followed the negotiated agreement approach~\cite{campbell2013coding} to discuss and resolve any disagreement, therefore improving the reliability of the analysis results for RQ3 and RQ4. In the final analysis, we generated four characteristics of architecture related questions that have more than one answer and five characteristics of architecture related questions that are that only have one answer as the results of RQ3 and RQ4, respectively. The details about these four characteristics for RQ3 and five characteristics for RQ4 are provided in Section \ref{resultsofRQ3} and Section \ref{resultsofRQ4}.

\textit{b) For analyzing RQ2}

We employed pre-defined classifications in \cite{bedjeti2017modeling} and \cite{petersen2009context} to answer RQ2. Specifically, the first author manually analyzed the extracted data for RQ2 (i.e., design contexts, see Table \ref{dataExtraction}) from the questions in the ARPs relevant to answering RQ2 (see Figure \ref{DrawingStudyExcution}). The first author then examined the extracted data to investigate the design contexts in which architecture related questions were raised. By referring to the categories of design contexts presented in the abovementioned studies, three main categories and eight subcategories were generated from the analyzed ARP questions. The personal bias was mitigated through the validation of the generated categories and subcategories with the other three authors of this study. As in the analysis of the previous RQs, the disagreements were discussed and resolved in a meeting using the negotiated agreement approach~\cite{campbell2013coding} to improve the reliability of the analysis results for RQ2. The results of RQ2 are presented in Section \ref{resultsofRQ2}.

\textit{c) For analyzing RQ5 and RQ6}
 
These two RQs aim to investigate the usefulness of architecture solutions in SO. As discussed in Section \ref{PhaseII_sampleSize}, in order to answer these two RQs, we defined a set of criteria (see Table \ref{subInclusionExclusion}) and filtered ARPs with useful knowledge. For clarity, we examined these two RQs from two aspects: 

We investigated how SO users discuss the usefulness of architecture solutions attributed to their associated architecture related questions. We needed to gain insights into ways (e.g., terms) SO users may use to communicate the usefulness of architecture solutions in SO. In addition, understanding SO users’ discussions on the usefulness of these solutions is important to direct Q\&A platform owners in creating the mechanisms that can help their users to efficiently and effectively search and (re)use such useful architecture solutions. To achieve this, we manually checked comments attached to the solutions in the 968 ARPs with the aid of our defined criteria (see Table \ref{subInclusionExclusion}). We found that SO users occasionally use terms related to usefulness, such as “helpful”, in the comments along with other terms (e.g., “very”, “super”) to explicitly convey how useful they found certain architecture solutions (e.g., see Figure \ref{UsefulComment}). As stated in Section \ref{datacollection}, the number of resulting ARPs (with useful knowledge) that were used to answer the two RQs (RQ5 and RQ6) is 324 ARPs (see Figure \ref{DrawingStudyExcution}). It is worth noting that we did not count on the occurrence of the terms, e.g., “useful” (and similar) stated in comments to measure the usefulness of such architecture solution given to an architecture related question in our study. As explained above, we referred to information in the comments attacked to the solutions to examine SO users' discussions on the usefulness of architecture solutions. We mean the reaction of SO users after seeing and using architecture solutions given to their associated architecture related questions, for example, see a comment in Figure \ref{UsefulComment}. Moreover, before such ARPs (solutions) were finally included for analysis, we first read the solutions commented to be helpful and their associated questions to check if they are really useful (i.e., are the solutions/answers useful to address the questions? \cite{zhu2009multi}). In Section \ref{resultsofRQ5}, we provide more details about the identified terms related to usefulness (e.g., “helpful”) along with other terms (e.g., “extremely”, “very”) (from 324 ARPs) that SO users use in comments to explicitly signify how useful they found architecture solutions provided to their associated questions.

During the data analysis for these two RQs (i.e., the taxonomy of architecture solutions considered useful (RQ5) and their characteristics (RQ6)), the first author followed the same procedures (e.g., coding and grouping similar codes into high-level categories) that were used when analyzing RQ1. One thing to elucidate when analyzing RQ5 to construct a taxonomy is that the first round of grouping yielded seven main categories. In the second round, the all the authors of this study proceeded to further generate subcategories and types from these seven main categories, ensuring that these main categories, their subcategories, and types follow an “\textit{is a}” relationship. The grouping process was iterative, in which the authors continuously went back and forth between categories, subcategories, types, and solutions to refine the taxonomy. The final results of this analysis yielded a taxonomy of 7 main categories, 20 subcategories of which 1 were encoded as “Others” (i.e., refer to codes that did not fit into the already generated subcategories), and 85 types. Note that the negotiated agreement approach \cite{campbell2013coding} was used to discuss and resolve any disagreements. The final taxonomy as the result of RQ5 is elaborated in Section \ref{resultsofRQ5}. Four characteristics were distilled as the results of RQ6 and are detailed in Sections \ref{resultsofRQ6}. 

Note that while categorizing and characterizing ARPs in SO by reading through those posts, we observed that a single ARP may contain multiple types of architecture knowledge. For example, in this ARP\footnote{\url{https://tinyurl.com/2d8r6w8m}} from our dataset, an SO user asked about alternative architecture patterns for Model View Controller (MVC) pattern (i.e., alternative architecture solutions). In the question body, the user asked the reasons that could drive someone to decide to use those alternative architecture patterns over MVC (i.e., architecture decisions and their rationale), the types of systems that the alternative architecture patterns are typical used for (i.e., design context), and the pros and cons that come along with using those alternative architecture patterns (i.e., benefits and drawbacks of architecture solutions). We encoded such a post with multiple types of architecture knowledge accordingly. Moreover, while analyzing ARPs in our dataset, we noted down and then discussed the results (e.g., categories of architecture related questions and taxonomy of useful architecture solutions) during the qualitative data analysis. This has led to several interesting findings and actionable implications for various stakeholders, which are presented in Section \ref{results} and Section \ref{discussionAndImplications}, respectively. The dataset collected and used in this study and the details of data analysis (e.g., coding in MAXQDA) are available online for replication and validation purposes \cite{dataset}.

\section {Results} \label{results}

In this section, we present the results to our RQs that we got from data analysis (see Section \ref{dataAnalyis}). The result of each RQ is presented in a dedicated subsection, ending with the key findings of the corresponding results.

\subsection{Categories of architecture related questions (RQ1)} \label{resultsofRQ1}
Categories of architecture related questions that SO users ask in SO were determined using the open coding and constant comparison techniques described in Section \ref{dataExtractionAndAnaly}. We examined questions in the 968 ARPs to answer this RQ (see Figure \ref{DrawingStudyExcution}). Our data analysis yielded 9 main categories and 21 subcategories of architecture related questions. Table \ref{CategoriesOfQuestions} shows the mentioned categories, their subcategories, their percentages of occurrence (out of 968 ARP questions), and count information. As shown in Table \ref{CategoriesOfQuestions}, \textit{architecture configuration} (27\%, 261 out of 968 ARP questions), \textit{architecture decision} (19\%, 181 out of 968 ARP questions), and \textit{architecture concept} (15\%, 142 out of 968 ARP questions) are the top three categories of most frequently asked architecture related questions. In the following, we report those categories and subcategories. Where required, we provide an SO question example to support the understanding of the categories and their subcategories.

\begin{table}[h]
\small
\centering
\caption{Categories of architecture related questions, their subcategories, and their counts \& percentages}
\label{CategoriesOfQuestions}
\resizebox{\textwidth}{!}{%
\begin{tabular}{lll} 
\toprule
                 Category                                 & Subcategory                                                 & Count \\ \hline

\multirow{2}{*}{Architecture configuration (27\%, 261)}       
                                                          & Architecture configuration with technologies support        & 144  \\ \cline{2-3} 
                                                          & Architecture pattern configuration                          & 117  \\ \hline
                                      
\multirow{2}{*}{Architecture decision (19\%, 181)}       
                                                          & Technology decision                                         & 104  \\ \cline{2-3} 
                                                          & Behavioral decision                                         & 77   \\ \hline
                                      
\multirow{4}{*}{Architecture concept (15\%, 142)}         & Architecture overview                                       & 62   \\ \cline{2-3} 
                                                          & Basic architectural concept                                 & 32   \\ \cline{2-3} 
                                                          & Architecture component functionality                        & 26   \\ \cline{2-3} 
                                                          & Specific architecture pattern                               & 22   \\ \hline
                                      
\multirow{2}{*}{Architecture implementation (12\%, 119)}       
                                                          & Architecture component implementation                       & 79  \\ \cline{2-3} 
                                                          & Architecture pattern implementation                         & 40  \\ \hline                                      
                  
\multirow{3}{*}{Architecture tool (10\%, 99)}& Architecture modeling tool                                               & 34  \\ \cline{2-3} 
                                                          & Model-based code generation tool                            & 30  \\ \cline{2-3} 
                                                          & Usage of architecture tool                                  & 21  \\ \cline{2-3} 
                                                          & Code-based model generation tool                            & 14  \\ \hline
                                      
\multirow{2}{*}{Architecture evolution (6\%, 55)}         & Architecture extension to meet new requirements             & 42  \\ \cline{2-3} 
                                                          & Component extension to meet new requirements                & 13  \\ \hline
                                      
\multirow{3}{*}{Architecture refactoring (5\%, 45)}       & Refactoring of circular dependencies                        & 21  \\ \cline{2-3} 
                                                          & Refactoring of large components                             & 13  \\ \cline{2-3} 
                                                          & Refactoring of big ball of mud                              & 11  \\ \hline
                                      
\multirow{2}{*}{Architecture deployment (4\%, 34)}        & Application  deployment to meet quality attributes          & 24  \\ \cline{2-3} 
                                                          & Application  deployment to meet functional requirements     & 10  \\ \hline
                                      
\multicolumn{2}{l}{Architecture documentation (3\%, 32)}                                                                & 32  \\ 

\bottomrule 
\end{tabular}%
}
\end{table}

\textbf{(1) Architecture configuration} questions in this category ask about how to configure components and connectors in software systems. The types of components and connectors could either belong to certain technologies (e.g., Windows Communication Foundation (WCF) and Windows Presentation Foundation (WPF)) or other architectural concepts (e.g., architecture patterns). This category is the most common (27\%, 261 out of 968 ARP questions) category of architecture related questions in SO (see Table \ref{CategoriesOfQuestions}). We further classified this category into two subcategories, in which \textit{architecture configuration with technologies support} surpasses half of the questions (144 out of 261 ARP questions of the \textit{architecture configuration} category) that SO users ask in this category (see Table \ref{CategoriesOfQuestions}). 

\begin{itemize} 
 
\item\textit{Architecture configuration with technologies support} is concerned about how to configure an architecture of an application with specific technologies (e.g., WPF, WCF). For instance, in this question\footnote{\url{https://tinyurl.com/yuxjp2su}}, a developer asked about how to configure or build a scalable Web-based application by using WCF: “\textit{Does anyone have any experience with how well web services build with Microsoft's WCF will scale to a large number of users? The level I'm thinking of is in the region of 1000+ client users connecting to a collection of WCF services providing the business logic for our application (...)}”.

\item\textit{Architecture pattern configuration} seeks practical guidance on how to configure a specific architecture pattern (e.g., Model View Controller and hexagonal architecture patterns) when designing an application to achieve certain requirements (e.g., functional requirements). For example, in this question\footnote{\url{https://tinyurl.com/4kn6t27e}}, a developer asked about how to configure an application that conforms to a Hexagonal architecture pattern by stating that: “\textit{I'm looking for some guidance or best practices for how to configure and structure an application which conforms to Hexagonal architecture that supports multiple (driver) adapters simultaneously (...)}”.
\end{itemize}

\textbf{(2) Architecture decision}: SO users ask this type of questions mostly when they want to decide between two or more alternative architecture solutions when deigning their software systems. Among two subcategories identified in this category, the \textit{technology decision} subcategory contains the majority of questions (104 out of 181 ARP questions of the \textit{architecture decision} category) that SO users ask in this category (see Table \ref{CategoriesOfQuestions}).

\begin{itemize} 
\item\textit{Technology decision} is mainly concerned about choosing between two or more technology solutions (e.g., frameworks, databases) to meet certain requirements at the architecture level. Moreover, various aspects can be considered during this choice, such as technology features, benefits, and drawbacks \cite{soliman2015enriching}. For instance, in this question\footnote{\url{https://tinyurl.com/k9xzkman}}, a developer asked about the reasons that could drive him or her to decide to use Cassandra over HBase for his/her application by stating that: “\textit{HBase is known for being a key-value store and random reads with .get and .put functions based on the key. Is Cassandra a better choice for suiting a requirement of key-value store? Can it support random reads based on key? If so, in which conditions should I choose Cassandra over HBase in a Spark Streaming application?}”.

\item\textit{Behavioral decision} is concerned with deciding how certain elements in a system would interact together to provide some functionality or to satisfy certain quality attributes \cite{kruchten2004ontology}. For example, a developer wanted to decide on either to let clients connect directly to the database or let the connection go through the web service by asking this question\footnote{\url{https://tinyurl.com/4pjh3ufk}}: “\textit{Recently I have been developing a system to run a high secured database (using vb.net and SQL Server 2005). I want to increase the security of the database so no connection will be made directly to the database but instead a HttpWebRequest is sent to a web service which then connects to the database and returns the requested data table in XML format. My concern is just about the performance, I cannot decide either to let clients connect directly to the database or let the connection go through the web service}”.

\end{itemize}

\textbf{(3) Architectural concept} includes theoretical related questions about software architecture. We divided this category into four subcategories, among which \textit{architecture overview} contains the majority of questions (62 out of 142 ARP questions of the \textit{architecture concept} category) that SO users ask in this category. 

\begin{itemize} 

\item\textit{Architecture overview} questions are concerned with the information about the general working mechanism or overview of certain existing architecture. For instance, in this question\footnote{\url{https://tinyurl.com/2a9hb3ek}}, a developer asked about architectural overview of Drupal version 7: “\textit{Could someone provide an architectural overview of the Drupal 7 control flow? Perhaps in the sense of a flowchart about how a page gets generated (...)}”.

\item\textit{Basic architectural concept} refers to questions that seek explanations about basic concepts in software architecture. For instance, in this question\footnote{\url{https://tinyurl.com/5burxuca}}, a developer was seeking explanations about several architecture concepts, such as a architecture pattern: “\textit{Is MVC a pattern or architecture or framework? What is a pattern? What is an architecture? (...)}”.

\item\textit{Architecture component functionality} is concerned with the use, purpose, or functionality of certain components in the architecture. For example, in this question\footnote{\url{https://tinyurl.com/3x35jzu6}}, a developer was asking about the use or purpose of the lifecycle aware component in Android based application: “\textit{We already have a Lifecycle in our Activity/Fragment then why will we use Lifecycle aware component \& kindly guide me the main purpose of it. And if we use lifecycle aware then why we use lifecycle that we knew already}”.

\item\textit{Specific architecture pattern} questions ask about particular architecture patterns that are commonly used in the design of certain applications to address functional or non-functional requirements. For instance, in this question\footnote{\url{https://tinyurl.com/2u2df8z5}}, a developer asked about commonly used architecture patterns for three-dimensional (3D) video game applications: “\textit{What are some of the more common design patterns used when developing 3D games? Are there any high-level architectural design patterns that are commonly used? (...)}”.
\end{itemize}

\textbf{(4) Architecture implementation} questions ask about how to implement a certain software system according to its architecture design. The architecture design is refined in detailed design, and then implemented in code. Architecture implementation category has two subcategories, among which \textit{architecture component implementation} occupies the majority of questions (79 out 119 ARP questions of the \textit{architecture implementation} category) that SO users ask in this category. 

\begin{itemize} 
\item\textit{Architecture component implementation} is concerned with how components should be implemented in the system. For instance, in this question\footnote{\url{https://tinyurl.com/kp73y3wk}}: “\textit{How to implement a single component sharing in different modules in Angular 7 while using lazy loading?}”.

\item\textit{Architecture pattern implementation} questions are about the ways certain architecture patterns are implemented with regard to the fundamental design principles. For example, in this question\footnote{\url{https://tinyurl.com/mpmvwb5c}}: “\textit{How to implement MVC in Swift? I've been building Swift apps where basically all the functionality is in the ViewController. I know this isn't the optimal way to do it because design patterns help you expand the app but I don't really understand them (...). How do I go about turning this into a Model-View-Controller design?}”.
\end{itemize}

\textbf{(5) Architecture tool}: There are various architecture tools (e.g., Enterprise Architect, Archi, Cloudcraft) that can be used to assist in the architecture design of a software system. With our dataset, we found architecture related questions in which SO users ask about these tools and classified them in the \textit{architecture tool} category. We further classified this category into four subcategories, among which \textit{architecture modeling tool} contains the majority of questions (34 out of 99 ARP questions of the \textit{architecture tool} category) that SO users ask in this category.

\begin{itemize} 
\item\textit{Architecture modeling tool} questions ask about tools that can enable the creation or drawing of architectural diagrams to model or represent an architecture of a software system during the design. For example, in this question\footnote{\url{https://tinyurl.com/ndf7mrnc}}: “\textit{I am a newbie in TOGAF and I need to start a first trial. I am trying to model my architecture. Which tool do you advise me to use in order to model my architecture using TOGAF?}”.

\item\textit{Model-based code generation tool} refers to questions that ask about architecture tools that can enable the generation of code from architectural models. For instance, in this question\footnote{\url{https://tinyurl.com/jwkvtzwc}}: “\textit{Please suggest me any open source tool to generate C\# code from UML designer (...). My requirement is to have a code generation tool for C\#}”.

\item\textit{Usage of architecture tool} questions look for instructions on how to use certain architecture tools (e.g., Archi, Microsoft Visio). For instance, in this question\footnote{\url{https://tinyurl.com/52zffbw3}}: “\textit{How to add UML/layer diagram to an existing solution in VS 2015 community? There is no architecture menu there?}”.

\item\textit{Code-based model generation tool} is concerned with tools that assist in architectural models generation or recovery from the codebase. For example, in this question\footnote{\url{https://tinyurl.com/n7rz24mu}}: “\textit{I need to make a UML class diagram for a project (...) I do not really want to write all the classes/functions manually, so I was trying to generate the diagram from the source code but can't seem to find a way or tool to do it. (...)}”. 
\end{itemize}

\textbf{(6) Architecture evolution}: SO users ask this type of architecture related questions when seeking help on how they can re-architect and expand their existing architecture for the purposes of achieving certain new requirements (functional or non-functional requirements). Among two subcategories identified in this category, the \textit{architecture extension to meet new requirements} subcategory contains the majority of questions (42 out of 55 ARP questions of the \textit{architecture evolution} category) that SO users ask in this category (see Table \ref{CategoriesOfQuestions}).

\begin{itemize} 
\item\textit{Architecture extension to meet new requirements} is concerned with practical guidance for expanding an existing architecture of a system to address certain new functional or non-functional requirements. The changes do not only happen in one component, but they may happen in almost the whole architecture of the system. For example, in this question\footnote{\url{https://tinyurl.com/ymbyzcvz}}: “\textit{I am expanding/converting a legacy Web Forms application into a totally new MVC application. The expansion is both in terms of technology as well as business use case (...). The new project has two primary goals: Extensibility (for currently and future pipeline requirements) and Performance (...). Is there a way in DDD to achieve both, Extensibility that DDD provides and performance that DBDD provides?}”.

\item\textit{Component extension to meet new requirements} includes questions that ask about the extension of certain architectural components to meet some new functional or non-functional requirements in existing and running software systems. This is different from the above subcategory, as here the change or extension happens in local to a specific component or layer, rather than affecting the whole architecture of the system. For instance, in this question\footnote{\url{https://tinyurl.com/3rh9xmhs}}: “\textit{We are currently evaluating CQRS and Event Sourcing architecture (...). What happens if, after an application has been up and running for a while, there is a new requirement to add an additional field to a ViewModel on the ReadModel database? Say, the Customer Zip Code is required on the CustomerList ViewModel, where it was not previously}”.
\end{itemize}

\textbf{(7) Architecture refactoring}: SO users ask this type of architecture related questions when they want to restructure architecture of systems aiming at improving non-functional attributes of those systems without modifying their external behaviors. This category includes three subcategories, where most of the questions (21 out of 45 ARP questions) are related to the subcategory \textit{refactoring of circular dependencies}.

\begin{itemize}
\item\textit{Refactoring of circular dependencies} is concerned with techniques and tools that can help remove undesirable circular or cyclic dependency issues among modules so that layering violations can be addressed and dependency structure can be improved in the systems. For instance, in this question\footnote{\url{https://tinyurl.com/475dvbp5}}: “\textit{I am working on the MVC project where I am following the layered architecture (...). Now, my Business Logic Layer(BLL) is depending on the Data Access Layer (DAL) which is depending on BLL because domain objects are inside BLL. So, both are having reference to each other (...). How can I overcome the circular dependency?}”.

\item\textit{Refactoring of large components} is concerned with approaches that can help refactor large components in software systems. For instance, in this question\footnote{\url{https://tinyurl.com/475dvbp5}}: “\textit{I have a pretty large table component and I want to separate its body section into new component. Each time I am trying to do this, the styling of table gets broken (...). I would like to have exactly this same page after this refactoring. Does anyone know how to pass styling to this new child component, or how to make thing styling work again ?}”.

\item\textit{Refactoring of big ball of mud}: Big ball of mud occurs when a software system lacks a perceivable, flexible, and appropriate architecture \cite{foote1997big}. This subcategory includes questions that ask about approaches and tools for big ball of mud refactoring. For instance, in this question\footnote{\url{https://tinyurl.com/j6rdefeb}}: “\textit{What step would you take to refactor a ball of mud CF app into something modern and maintainable}”.
\end{itemize}

\textbf{(8) Architecture deployment} collects architecture related questions that ask about how certain software systems should be deployed in the hosting environments to meet requirements (e.g., functional and non-functional requirements). According to our studied dataset, we divided this category into two subcategories, among which \textit{application deployment to meet quality attributes} contains the majority of questions (24 out of 34 ARP questions of the \textit{architecture deployment} category) that SO users ask in this category.

\begin{itemize} 
\item\textit{Application deployment to meet quality attributes} includes architecture related questions that ask about methods and tools that assist in the deployment of applications in the hosting environments to meet quality attributes (e.g., availability and performance). For instance, in this question\footnote{\url{https://tinyurl.com/4tyd4yt6}}, a developer asked how to deploy a microservice based system with zero downtime: “\textit{At the moment I'm working on an application which will be based on the Microservice architecture. As main technologies, we planned to use Spring Boot and Docker for each Micro Service development. One of the goals/requirements is to provide a Zero Downtime Deployment feature for the users (...). Any suggestions on the Zero Downtime Deployment process? If you have any great ideas for a different architecture or maybe you’ve used tools which can help us here (...)}”.

\item\textit{Application deployment to meet functional requirements} covers architecture related questions that ask about methods and tools that assist in the deployment of software systems to meet functional requirements. For example, in this question\footnote{\url{https://tinyurl.com/37dmd6av}}, a developer asked about the method s/he can follow in order to deploy his/her microservices based application in the production environment so that each service of the application can call each other: “\textit{I am trying to deploy my microservices architecture to production env. Now I have 15 services, 1 Facade Layer, Facade Layer calls services, gets data, aggregates them, and generates the final result. Also, services call each other(rarely but yes, they call each other) (...). So I have decided that I will have 5 Boxes (5 high-end servers). A, B, C, D, E A will be LVS (for Load Balancing) B \& C will host the Facade layer. So when the request came for Facade, it will come from A and load balanced to B \& C (...). So B \& C box will contain each one haproxy instance also since when Facade Layer calls services, it will be load balanced (...). But my question is how should I allow my services to call each other? (...)}”.
\end{itemize}

\textbf{(9) Architecture documentation}: This is the only category of architecture related questions with no subcategories in our studies dataset. The \textit{architecture documentation} category includes questions that ask about methods and tools that assist in the documentation of architecture of software systems. For instance, in this question\footnote{\url{https://tinyurl.com/msnbc7xb}}, a developer asked about the best practices and tools for documenting architecture of different types of systems: “\textit{What are the best practices and software tools for documenting software design and architecture for PC based applications based on Java or .NET? Embedded Applications based on VxWorks or Embedded Linux or Windows CE? (...)}”.

\begin{tcolorbox}[colback=gray!5!white,colframe=gray!75!black,title=Key Findings of RQ1]
\textbf{Finding 1}:
SO users ask a broad range (9 categories) of architecture related questions, among which \textit{architecture configuration} (27\%, 261 out of 968 ARP questions), \textit{architecture decision} (19\%, 181 out of 968 ARP questions), and \textit{architecture concept} (15\%, 142 out of 968 ARP questions) are the top three categories of most frequently asked architecture related questions. 
\end{tcolorbox}

\subsection{Categories of design contexts (RQ2)} \label{resultsofRQ2}

This RQ aims to investigate the categories of design contexts in which architecture related questions were raised. As described in Section \ref{dataExtractionAndAnaly}, to answer this RQ, we used a predefined classifications of design contexts from \cite{bedjeti2017modeling} and \cite{petersen2009context}) when analyzing the extracted data for RQ2 (i.e., design contexts) from the 968 ARP questions (see Figure \ref{DrawingStudyExcution}). We found that most (71\%, 687 out of 968) of our analyzed ARP questions describe their design contexts (i.e., the knowledge about the environments in which the systems are expected to operate \cite{bedjeti2017modeling}), and then the responders provided potential solutions with rationale based on the given design issues and design contexts. In addition, we identified three main categories and eight subcategories of design contexts. We report the mentioned categories, their subcategories, their percentages of occurrence (out of 687 ARP questions), and count information in Table \ref{summaryOfDesignContexts}. It is also evident from Table \ref{summaryOfDesignContexts} that \textit{application context} is the most common (54\%, 377 out of 687 ARP questions) category of design contexts, and \textit{organizational context} is the least significant category (8\%, 56 out of 687 ARP questions).

\begin{table}[h!]
\small
\centering
\caption{Categories of design contexts, their subcategories, and their counts \& percentages}
\label{summaryOfDesignContexts}

\begin{tabular}{p{5cm}p{6cm}p{1cm}}
\toprule
                Design Context                          & Subcategory                       & Count  \\ \hline
\multirow{2}{*}{Application context     (55\%, 377)}    & Application domain context        &  313   \\ \cline{2-3} 
                                                        & External service context          &   64   \\ \hline

\multirow{2}{*}{Platform context        (37\%, 254)}    & Software context                  &  139   \\ \cline{2-3} 
                                                        & Hardware context                  &  115    \\ \hline                                     
                                                     
\multirow{3}{*}{Organizational context  (8\%, 56)}      & Development schedule context      &   36   \\ \cline{2-3} 
                                                        & Stakeholders context              &   13   \\ \cline{2-3}
                                                        & Resources context                 &    7   \\           
\bottomrule 
\end{tabular}%
\end{table}

\textbf{(1) Application context} refers to the software system or product that is to be designed. It is accessed through a device (platform entity) to deliver services to end-users \cite{bedjeti2017modeling}. This category includes two subcategories, in which the \textit{application domain context} subcategory is the most (313 out of 377 ARP questions) common one.

\begin{itemize} 
\item\textit{Application domain context} describes the domain/type of the application that is being developed (such as E-commerce system, banking system, distributed system) \cite{petersen2009context}. Some SO users like to reveal in their architecture related questions what kind of application domains they are about to design in order to get potential and relevant architecture solutions that fit their application domains. For example, in this question\footnote{\url{https://tinyurl.com/2p93nesu}}, a developer mentioned that s/he was designing an E-commerce system: “\textit{I am designing an E-commerce using microservices architecture. Suppose that I have two contexts: a product catalog, inventory and pricing. It's seems clear to me that they have a clear responsibility. But to serve the show case (the product list) I need to make a request for the product catalog, get a list of ID's and then use it to query the Inventory micro services to check inventory status (in stock or stock out). Besides that I need to make a request to Pricing to get the price of each product (...). I have been reading about microservices architecture and when you are dealing with many `joins’ it's possible that the these contexts should be a single one (...). We can use a domain event to notify `search’ microsecond that something has changed. So we can resolve show case with a single request. This look like a CQRS. Is there a correct approach? Which one is better ? Trade-offs?}”. 

\item\textit{External service context} refers to specifications of external software services that the application uses \cite{bedjeti2017modeling}. For example, in this question\footnote{\url{https://tinyurl.com/2p8phmr6}}, a developer mentioned that s/he was designing a system that will require to use Azure or Amazon cloud services: “\textit{Basically my question is on the application architecture. Designing for hosting is easy but cloud computing adds new challenges (...). I am not certain what I should do in designing an application for safety engineers, so a high uptime is important. So, if my application is written in ASP.NET, using SQL Server, it would seem that my best bet is to design for Azure, but would Amazon's solution be a good choice? How would I decide if I should just have everything on the same system or have the data on Amazon's cloud and the ASP.NET on Azure? (...). I decide on the language, does that lock me into a cloud solution?}”.
\end{itemize}

\textbf{(2) Platform context} comprises the hardware technology a user employs to access an application, the software it runs, and the network capabilities of such technology \cite{bedjeti2017modeling}. In our dataset, we identified two platform contexts (i.e., software and hardware context). 

\begin{itemize} 
\item\textit{Software context} comprises information about the software elements of the device, such as the Operating System (OS) or other installed applications. This subcategory collects the ARP questions that describe the software elements of the device (e.g., OS) on which the planned software system will need to run in production \cite{neto2005toward}. We found that some SO users provide this kind of information when asking architecture related questions. For example, in this question\footnote{\url{https://tinyurl.com/2p8nd2kw}}: “\textit{I need to build one mobile application starts with \textbf{windows phone 7} and then need to convert the application to other platforms like \textbf{Android}, \textbf{iOS}. The application contains many screens with data capture and all the data stores it in local storage and finally, it is passed to a central server. I would like to know how the architecture needs to be designed (...)}”.

\item\textit{Hardware context} comprises the platform entity which defines the device through which the user accesses and uses the application, and can be of different types, such as desktop, laptop computers, and wearable mobile devices \cite{bedjeti2017modeling}. The hardware context category gathers ARP questions that mention hardware technologies (e.g., desktop computers) through which the users access and utilize the planned applications. For example, in this question\footnote{\url{https://tinyurl.com/yc2eyvhs}}, a developer mentioned that s/he was developing a desktop application: “\textit{We want to start develop an intermediate \textbf{desktop} software. We decided to use the WPF. We don't want to use the MVVM pattern. Because we are not familiar with MVVM. Is it true to develop WPF application without MVVM pattern (using 3 layer architecture but without MVVM) although does it have better performance than win forms yet?}”. 
\end{itemize} 

\textbf{(3) Organizational context} refers to the development schedule (e.g., time-to-market), the people (i.e., stakeholders), or the resources that could influence the development of software systems. This category includes two subcategories, in which the \textit{development schedule context} subcategory is the most (36 out of 56 ARP questions) common one.

\begin{itemize}
\item\textit{Development schedule} describes the time put on software development. For example, in this question\footnote{\url{https://tinyurl.com/2bu5yu65}} a developer mentioned that the development time for a software project was restricted to only three months: “\textit{For personal and university research reasons I am thinking of building a simple CRM using a service-oriented architecture (...). The architecture that I'm designing defines: - WebGUI (a client of the other services) - AnalyticsService (a service that receives data, analyzes, and collects it) - CustomerCareService (a service that uses RESTful APIs to apply CRUD operations (...). What sort of authentication is more suitable for a client (user token vs OAuth or similar). \textbf{I've about 3 months} to do it (...)}”.

\item\textit{Stakeholders context} describes the people who are involved in the development of a software system, for example, project managers, owners, architects, developers, users, among others. For instance, in this question\footnote{\url{https://tinyurl.com/yra7d3y7}} an asker mentioned a number of developers that was involved in the development of 3D Map application by stating that: “\textit{I'm trying to develop 3D Map, and I found 3 solutions: Use game engine (like unity) or Use 3D graphic API (OpenGL, etc) or Web app. Is there another way to do it? And which one of those three solutions (design decision) is better? (with reason) (..) \textbf{Developers: 3 programmers}}”.

\item\textit{Resources context} denotes the lack (or availability) of resources (e.g., financial or technological competencies) at disposal to develop an application \cite{petrov2011need}. For instance, one developer needed to update an application with a tight budget and stated in this question\footnote{\url{https://tinyurl.com/4pjp3ntj}} that: “\textit{I have to build a database/image-rich application that's only going to increase in size (scalability). I am on a budget, but do have a rather good 3Ghz Xeon server with 400 GB space. Any ideas? a good way for an individual on a \textbf{TIGHT budget}}”.
\end{itemize}

\begin{tcolorbox}[colback=gray!5!white,colframe=gray!75!black,title=Key Findings of RQ2]
\textbf{Finding 2}:
Most of the SO users (71\%, 687 out of 968 ARP questions) considered design contexts when asking architecture related questions.

\textbf{Finding 3}:
\textit{Application context} is the most common (54\%, 377 out of 687) category of design contexts in ARP questions, whereas \textit{organizational context} is the least significant design context category (8\%, 56 out of 687) in ARP questions.
\end{tcolorbox}
 
\subsection{Characteristics of architecture related questions that have more than one answer (RQ3)} \label{resultsofRQ3} 

\begin{figure}[H] 
 \centering
 \includegraphics[scale=0.7]{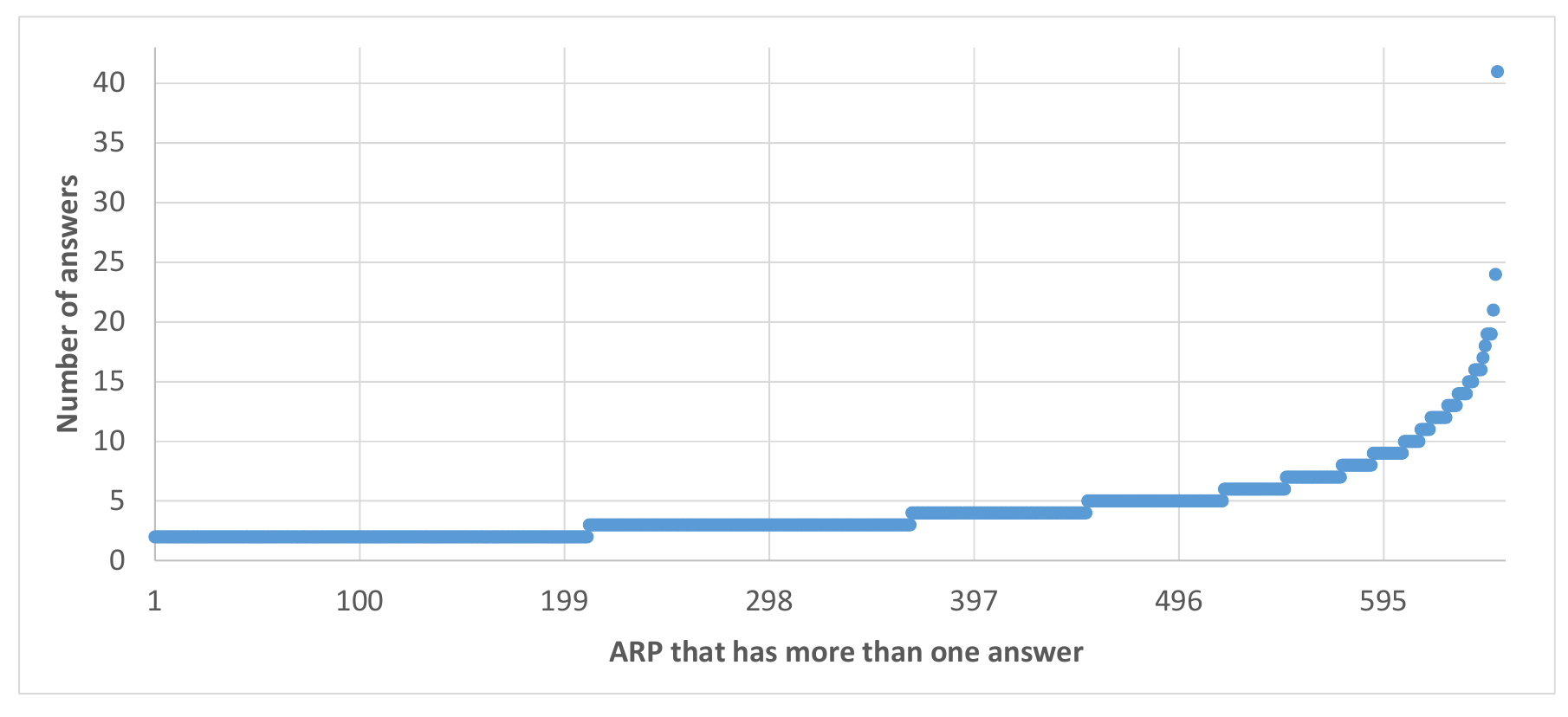}
 	\caption{Numbers of answers of the ARPs that have more than one answer}
 \label{AnswersDistributionOfARPquestionsWithMoreAnswers}
\end{figure}

Some architecture related questions are continuously getting more attention from SO users by answering them. This motivated us to investigate why such architecture related questions get more than one answer in SO by characterizing those questions. As mentioned in Section \ref{dataExtractionAndAnaly}, we used two techniques (i.e., open coding and constant comparison) from Grounded Theory \cite{stol2016grounded}, to examine the characteristics of ARP questions that have more than one answer from 650 ARPs (a subset of 968 ARPs) (see Figure \ref{DrawingStudyExcution}). As discussed in Section \ref{dataExtractionAndAnaly}, we referred to the contents of the questions and comments attached to questions to understand what factors (e.g., question formulation \cite{WritingGoodquest2018} or certain features in the question (e.g., architectural diagram)) that contribute to such architecture related questions getting more than one answer. The outputs of our data analysis generated four common characteristics of architecture related questions that have more than one answer. We provide these four common characteristics and their counts \& percentages in Table \ref{CharacteristicMoreAnswers}, which shows that \textit{well-articulated architectural information} is the most (46\%, 297 out of 650 ARP questions) frequent characteristic while \textit{upvoted architecture question} comes as the least (8\%, 51 out of 650 ARP questions) frequent characteristic of architecture related questions that have more than one answer. Moreover, we show the numbers of answers of the ARPs that have more than one answer in Figure~\ref{AnswersDistributionOfARPquestionsWithMoreAnswers}.

\textbf{(1) Well-articulated architectural information in the question}: 
The main reason an architecture related question would continuously be answered is that its architectural information is well-articulated. This question provides an overview of the planned software system and its basic principles (e.g., design contexts and architectural constraints) to help other SO users perceive what the question is really about (the purpose of the question). For example, a comment: “\textit{+1 great question, I want to say that this is a beautifully and very well-articulated question!)}” was posted under this question \footnote{\url{https://tinyurl.com/dfw27h38}} that illustrates and explains well the encountered architecture concerns (i.e., architecting a system to meet the scalability and visualization of data).

\textbf{(2) Clear description together with architectural diagrams in the question}: Another reason an architecture related question would continuously be answered is that it is easy to read and understand, for example, the question which clearly states necessary information about the components and connectors together with interfaces and relationships to other components. Also, providing diagrams, such as architectural component diagrams that depict and clarify the logical architecture view of software systems, contributes to questions continuously being answered. For example, before one responder started answering this question \footnote{\url{https://tinyurl.com/74fr8mwe}}, this responder stated that: “\textit{Your question is clearly described. Thanks for the little graph you drew to help clarify the overall architecture (...)}”.

\textbf{(3) Alternative architecture solutions to answer the question}: Although different architecture solutions act as alternative solutions to similar design problems, they differ in terms of their qualities \cite{soliman2015enriching}. For example, two architecture solutions may both address the interoperability concern but may differ into addressing performance concern. However, such alternative architecture solutions are of significant importance as they provide a wide range of possibilities for choosing and making design decisions on candidate architecture solutions for certain design issues. In this study, we noticed that questions that ask about choosing between various architecture solutions, such as technologies (e.g., databases, frameworks, and programming languages), are continuously getting new answers (alternative solutions). For example, SO users were interested in choosing a right combination of message formats with message transmission techniques in order to achieve quality attributes, including high performance, availability, scalability, among others, for their Ruby and Java applications interaction, and they posted a question\footnote{\url{https://tinyurl.com/44sey2a2}}: “\textit{We have cloud-hosted (RackSpace cloud) Ruby and Java apps that will interact as follows (....). We are interested in evaluating both message formatting (such as JSON) as well as message transmission techniques (RPC, REST, SOAP, etc.). Our criteria are high performance, availability, scalability (...). What combination of message format and transmission method would you recommend? Why?}”.

\textbf{(4) Upvoted architecture question}: Usually, in SO, if a question is consistently getting upvote count, its likelihood of being answered or getting more answers increases \cite{asaduzzaman2013answering}, and architecture related question is no exception. In our data analysis, we realized that this factor (upvoted architecture question) also contributes to architecture related questions having more than one answer. For example, in this architecture related post\footnote{\url{https://tinyurl.com/7bc8764a}}, a responder stated in the title of the answer thread when s/he was answering the question: “\textit{Since this question got upvoted several times I would like to share what I did in the end (...)}”.

 \begin{table} [h!] 
 \small
 \centering
	\caption{Characteristics of architecture related questions with more answers and their counts \& percentages}
	\label{CharacteristicMoreAnswers}
	\begin{tabular}{p{11cm}p{1.5cm}}
		\toprule
		Characteristic                                                                  & Count       \\
		\midrule
		Well-articulated architectural information                                      & 297 (46\%)  \\
		Clear description together with architectural diagrams in the question          & 174 (27\%)  \\
	    Alternative architecture solutions to answer the question                       & 118 (18\%)  \\
		Upvoted architecture question                                                   & 51   (8\%)    \\
		\bottomrule 
	\end{tabular}
\end{table}

 \begin{tcolorbox}[colback=gray!5!white,colframe=gray!75!black,title=Key Findings of RQ3]
\textbf{Finding 4}:
\textit{Well-articulated architectural information} is the most (46\%, 297 out of 650 ARP questions) frequent characteristic of architecture related questions that have more than one answer.

\textbf{Finding 5}:
The presence of architectural diagrams (e.g., components diagrams) in the architecture questions increases the chance of these questions to get more than one answer.
 \end{tcolorbox}

\subsection{Characteristics of architecture related questions that only have one answer (RQ4)} \label{resultsofRQ4}

We found out that some architecture related questions gain less attention from SO users to be continuously answered. Analogous to the two previous RQs (i.e., RQ1 and RQ3), we applied two techniques (open coding and constant comparison) to study the characteristics of questions that only have one answer from 318 ARPs (a subset of 968 ARPs) (see Figure \ref{DrawingStudyExcution}). As detailed in Section \ref{dataExtractionAndAnaly}, similar to RQ3, we referred to the contents of the questions and the comments attached to questions to understand what factors demotivate the responders to continue answering certain architecture related questions. We identified five common characteristics of those questions with their counts \& percentages in Table \ref{CharactertiscsFewAnswes}, which shows that \textit{lacking information in the question} (39\%, 125 out of 318 ARP questions) and \textit{poorly structured architecture question} (22\%, 69 out of 318 ARP questions) are the two major characteristics of architecture related questions that only have one answer. Below, we elaborate these characteristics in detail with examples from ARPs.

\textbf{(1) Lacking information in the question}: Architecture related questions that lack certain significant information (e.g., missing information on components and connectors together with interfaces and relationships to other components) fail to attract community members to provide their answers. For example, one developer pointed out that some information is missing in this architecture related question\footnote{\url{https://tinyurl.com/btukhpzx}}: “\textit{Your question cannot be answered without doing (many) assumptions. More information is needed about the module's dependencies. Are they stateless? Can you draw a flow of the requests?}”.

\textbf{(2) Poorly structured architecture question}: We found that architecture related questions that are poorly structured (e.g., not well articulated) fail to clearly reveal their purposes to the community members so that they can provide answers. These questions sound unclear, vague, or hard to follow. For example, in this question\footnote{\url{https://tinyurl.com/3m6fzjbe}} a developer asked about how to implement the interactors in Android MVP clean architecture but failed to clearly structure well his/her question, and another developer came and commented: “\textit{So what exactly is your question?}”. The asker came back and edited the question to make it clear so that other community members could understand what the question is really about. 

\textbf{(3) Architecture considered as off-topic}: Even though architecture related questions are being asked in SO, SO is mainly designed for programming information seekers. Therefore, architecture being not directly for programming related issues but mainly for high-level structure related concerns, this leads to the situation that architecture related questions get less attention from the community and consequently only get one answer or remain unanswered in SO. For example, a developer asked about the overview of ZeroMQ architecture, and another developer commented under this question\footnote{\url{https://tinyurl.com/rmpfarjc}} by saying that: “\textit{I'm voting to close this question as off-topic because it's not really about programming}”.

\textbf{(4) Proprietary technology in the question}: We found a few questions, asking about proprietary technologies, such as databases and frameworks that are not widely used, get less attention in SO and consequently get only one answer. For instance, in this architecture related post\footnote{\url{https://tinyurl.com/nrz66x3w}}, a community member claimed to be one of the technology founders of RethinkDB in the title of the answer when s/he was answering a question. Other community members kept asking more questions about that RethinkDB (a not widely used database) in the comment thread, and those questions have remained unanswered. For example, “\textit{Disclaimer: I'm one of the founders of RethinkDB. Sorry for the longish answer (...). RethinkDB is designed with a very flexible architecture (...)}”.

\textbf{(5) Duplicate architecture question}: Similar to other types of questions (e.g., programming questions) in SO, some architecture related questions get few answers (i.e., one answer) or remain unanswered because they are duplicate architecture questions. Community members do not like to re-answer questions that were answered before \cite{asaduzzaman2013answering}. They would like the askers to review the site (i.e., to check if their questions have not been posted and answered) before posting such new questions. For example, a comment: “\textit{duplicate of stackoverflow.com/questions/15142386/…}”, was posted under this architecture related question\footnote{\url{https://tinyurl.com/2p9382hh}}. 

\begin{table} [h!]
\small
 \centering
	\caption{Characteristics of architecture related questions with few answers and their counts \& percentages}
	\label{CharactertiscsFewAnswes}
	\begin{tabular}{p{9cm}p{1.5cm}}
		\toprule
		Characteristic                                                          & Count       \\
		\midrule
		Lacking information in the question                                     & 125 (39\%)  \\
		Poorly structured architecture question                                 & 69 (22\%)   \\
	    Architecture considered as off-topic                                    & 51 (16\%)   \\
		Proprietary technology in the question                                  & 32 (10\%)   \\
	    Duplicate architecture question                                         & 29 (9\%)    \\
		\bottomrule 
	\end{tabular}
\end{table}

 \begin{tcolorbox}[colback=gray!5!white,colframe=gray!75!black,title=Key Findings of RQ4]

\textbf{Finding 6}:
\textit{Lacking information in the question} (39\%, 125 out of 318 ARP questions) and \textit{poorly structured architecture question} (22\%, 69 out of 318 ARP questions) are the top two most frequent characteristics of architecture related questions that only have one answer.
 \end{tcolorbox}

\subsection{Taxonomy of architecture solutions that are considered useful (RQ5)} \label{resultsofRQ5}
\begin{landscape}
\begin{figure}
  \centering
  \includegraphics[scale=0.54]{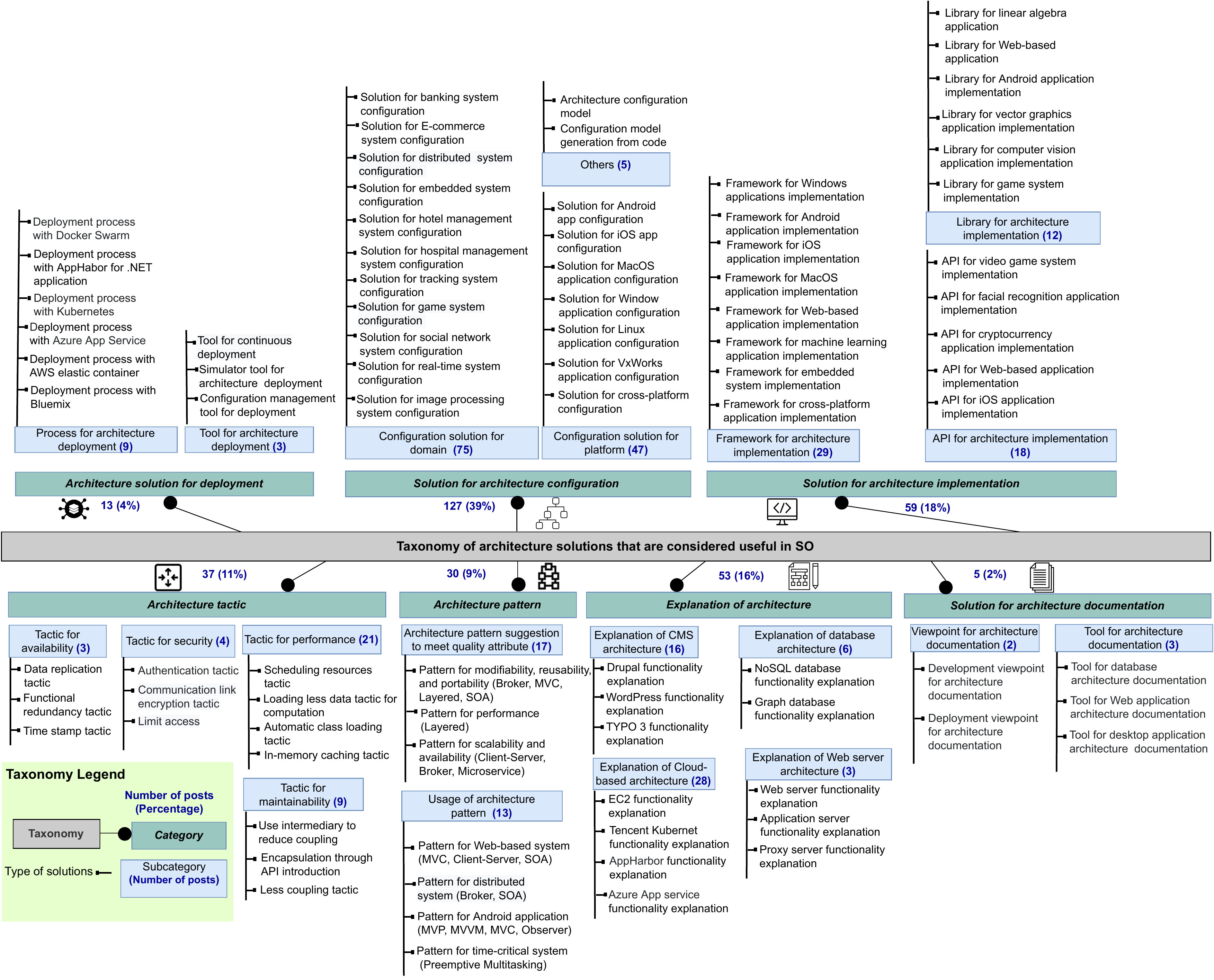}
 \caption{Taxonomy of architecture solutions that are considered useful}
 \label{TaxonomyOfArchSolution}
\end{figure} 
\end{landscape}

This RQ aims to construct a taxonomy of architecture solutions that are considered useful in SO. As discussed in Section \ref{dataExtractionAndAnaly}, when answering this RQ, we first investigated how SO users discuss the usefulness of architecture solutions attributed to their associated architecture related questions. We needed to gain an understanding of the ways (e.g., terms) SO users may use to communicate the usefulness of architecture solutions in SO. Understanding SO users’ discussions on the usefulness of these solutions is important to direct Q\&A platform owners in creating the mechanisms that can help SO users to efficiently and effectively search and (re)use such useful architecture solutions. We found that SO users frequently use two terms related to usefulness (i.e., “useful” and “helpful”) along with six other terms (i.e., “definitely”, “very”, “really”, “super”, “extremely”, and “incredibly”) in the comment threads (see Figure \ref{UsefulComment}) to explicitly express their feedback about how useful they found certain architecture solutions provided to their associated architecture related questions. Note that SO users may use other ways to communicate the usefulness of architecture solutions in SO, and we cannot claim that we have identified all usefulness terms. Secondly, as detailed in Section \ref{dataExtractionAndAnaly}, we thoroughly and comprehensively examined the contents of the solutions from 324 ARPs (a subset of 968 ARPs) with useful knowledge (see Figure \ref{DrawingStudyExcution}) to construct the taxonomy of these solutions. This examination yielded a taxonomy of 7 main categories, 20 subcategories of which 1 were encoded as “Others” (i.e., refer to codes that do not fit into the already generated subcategories), and 85 types (see Figure \ref{TaxonomyOfArchSolution}). 

\textbf{(1) Solution for architecture configuration}: This is the largest category of architecture solutions in our taxonomy (see Figure \ref{TaxonomyOfArchSolution}). The solutions in this category provide approaches and tools that enable the configuration of components and connectors of the planned software systems. Among three subcategories identified in this category, the \textit{configuration solution for domain} subcategory collects more than half of the solutions (75 out of 127 ARP solutions) discussed in this category (see Figure \ref{TaxonomyOfArchSolution}). 
\begin{itemize} 
\item\textbf{Configuration solution for domain}: This subcategory discusses approaches and tools for configuring applications of various domains, such as \textit{solution for distributed system configuration}, \textit{solution for banking system configuration}, \textit{solution for E-commerce system configuration}, and \textit{solution for game system configuration} (see Figure \ref{TaxonomyOfArchSolution}). Concerning the \textit{solution for distributed system configuration} type, a user asked about how to design and configure a 2/3 tier distributed application in Java with certain components, including centrally shared database and multiple fat clients (Swing based Graphical User Interface clients (GUIs)). S/he needed a simpler approach that could help him or her to configure those clients so that each client can be informed about data changes committed to the database by another client. The first solution in this ARP\footnote{\url{https://tinyurl.com/jfnuke2w}} is provided based on the application domain (i.e., \textit{distributed application}) described in the question. The solution suggests to configure the application's components (e.g., database and clients) by following Java EE distributed container/component-based architecture by stating that: “\textit{(...) Java EE is a distributed container/component based architecture for the enterprise tier (...) You c/would design a messaging domain with both topic/subscription based and straight up Queues. These can be declaratively configured to be durable, or not, etc. (...)}”.

\item\textbf{Configuration solution for platform} provides approaches and tools that enable the configuration of components and connectors of applications with regard to the platforms (e.g., Windows OS) on which these applications will run in production. We identified seven commonly discussed solution types in this category, such as \textit{solution for Android app configuration}, \textit{solution for Windows applications configuration}, \textit{solution for iOS app configuration}, and \textit{solution for cross-platform configuration} (see Figure \ref{TaxonomyOfArchSolution}). Regarding the \textit{solution for cross-platform configuration} type, one SO user needed to design and configure an application that sends data between two iOS devices (i.e., iPad and iPhone) with iPad acting as an iBeacon. The first solution in this ARP\footnote{\url{https://tinyurl.com/ytw52k6p}} explains how the application's components including the iPad and iPhone could be configured by using an approach that could support Android as well by stating that: “\textit{(...) I was forced into an architecture that would support Android as well, so I switched to BlueTooth. The iPad acting as an iBeacon also has BlueTooth code that is looking for ‘peripherals’ with a certain signature. Once the iPhone detects the iBeacon, the app then starts transmitting a BlueTooth peripheral signal with the appropriate signature (...)}”.
\end{itemize}

\textbf{(2) Solution for architecture implementation}: The ARP solutions in this category provide technology solutions, such as frameworks and libraries (see Figure \ref{TaxonomyOfArchSolution}), for implementing diverse architecture designs to address the system requirements (e.g., quality attributes). According to our dataset, we classified these technology solutions into three subcategories, among which \textit{framework for architecture implementation} contains the majority of solutions (21 out of 59 ARP solutions) that SO users discuss in this category (see Figure \ref{TaxonomyOfArchSolution}). 

\begin{itemize} 
\item\textit{Framework for architecture implementation}: These solutions gather different types of frameworks for implementing architecture design, for example, \textit{framework for Web-based application} (such as Laravel, Django, Express.js, and Play frameworks), \textit{framework for iOS application} (such as SwiftUI, Flutter, and React Native frameworks), and \textit{framework for Windows applications} (such as WinForm, WPF, and UWP frameworks) (see Figure \ref{TaxonomyOfArchSolution}). Regarding the \textit{framework for Web-based application} type, a user asked about (among other things) a framework that could facilitate the implementation of REST APIs in a Web-based application which will serve the content to mobile apps. The third answer in this ARP\footnote{\url{https://tinyurl.com/2p8pm9rs}} suggests the \textit{Play framework} as the solution to that question by stating that: “\textit{Use Play! to do it all. Writing REST services in Play is very very easy (...)}”.
 
\item\textit{API for architecture implementation} accumulates different types of APIs as solutions to questions that ask about APIs for implementing architecture design, for instance, \textit{API for video game system} (such as Pokeapi, Chicken Coop, Dota2, and Minecraft APIs), \textit{API for Web-based application} (such as REST, SOAP, RPC, and Geolocation APIs), and \textit{API for facial recognition application} (such as Lambda labs and Microsoft Computer Vision APIs) (see Figure \ref{TaxonomyOfArchSolution}). Concerning the \textit{API for Web-based application} type, one developer needed an API to implement a request-response Web application in Service Orientated Architecture (SOA) in order to meet certain requirements (including high performance). The second answer in this ARP\footnote{\url{https://tinyurl.com/pakzw2yk}} suggests to use \textit{REST API} over \textit{SOAP API} by noting that: “\textit{(...) consider also using REST API, it demands less overhead than SOAP, and you can use JSON as document format which is also more compact than XML, lowering network throughput requirements (...) SOAP has more fancy features that are not well supported in all languages, if you use REST you will be more safe here (...)}”.

\item\textit{Library for architecture implementation}: These solutions recommend various libraries for implementing architecture, for example, \textit{library for linear algebra application} (such as JBLSA, MTJ, OjAlgo, and EJML), \textit{library for computer vision application} (such as OpenCV library), \textit{library for vector graphics application} (such as DISLIN library), and \textit{library for Web-based application} (such as HPPC, Trove, and FastUtil) (see Figure \ref{TaxonomyOfArchSolution}). Regarding the \textit{library for vector graphic application} type, the first answer in this ARP\footnote{\url{https://tinyurl.com/259h4f6e}} suggests Raphaël library as the solutions to the question that asks about vector graphics application by saying that: “\textit{(...) I chose RaphaëlJS and I have to say it has been an absolute pleasure to use, and the help is fantastic too (...)}”.
\end{itemize}

\textbf{(3) Explanation of architecture}: The ARP solutions in this category provide theoretical explanations, purposes, or functionalities of architecture instead of providing concrete instructions on how to do something (e.g., how to configure certain architectural components in the system). Explanation of architecture category consists of four subcategories, among which \textit{explanation of cloud-based architecture} is the most (28 out 53 ARP solutions) discussed subcategory (see Figure \ref{TaxonomyOfArchSolution}). 

\begin{itemize} 
\item\textit{Explanation of cloud-based architecture} provides theoretical explanations, differences, and functionalities of the architecture of cloud computing services, such as \textit{Azure App service functionality explanation}, \textit{EC2 functionality explanation}, and \textit{AppHarbor functionality explanation} (see Figure \ref{TaxonomyOfArchSolution}). For instance, a user asked about the difference between \textit{Azure App Service} and the A\textit{Azure Service Fabric} in terms of functionalities in software development. The fourth answer in this ARP\footnote{\url{https://tinyurl.com/2p8hnkmm}} provides a detailed explanation about the difference between those two Azure platforms in terms of functionalities in software development as the solution to that question by stating that: “\textit{(...) They're two separate platforms, following different development paradigms. The App Service will give you functionality that Service Fabric doesn't provide out of the box. Stuff like auto-scale, authentication, rate limiting, integration with SaaS applications, etc. (...)}”. 

\item\textit{Explanation of CMS architecture} describes the functionalities of the architecture of Content Management Systems (CMS), such as \textit{Drupal functionality explanation} and \textit{TYPO 3 functionality explanation} (see Figure \ref{TaxonomyOfArchSolution}). For example, the first answer in this ARP\footnote{\url{https://tinyurl.com/2a9hb3ek}} provides the architecture overview of \textit{Drupal} (together with an architectural diagram) as the solution to the question that asks about the functionality of Drupal (e.g., control flow and how a page gets generated) by saying that: “\textit{(...) Although it's procedural PHP, it's purely event/listener driven in its architecture, and there's no simple ‘flow’ in the main PHP script for you to look though (...) Drupal's index.php file functions as a front-side controller (...)}”.

\item\textit{Explanation of database architecture} groups the ARP solutions that explain or describe the architecture of datab”ase systems, for instance, \textit{NoSQL database functionality explanation} (such as Apache Cassandra) and \textit{Graph database functionality explanation} (such as Nebula Graph) (see Figure \ref{TaxonomyOfArchSolution}). For example, the first answer in this ARP\footnote{\url{https://tinyurl.com/wpvw6j5h}} provides an explanation about \textit{ Cassandra} in terms of data replication to deal with data failure scenario as the solution to the question that asks about the way Cassandra handles such a scenario if one node goes down containing the record (data) a user is querying by stating that: “\textit{Cassandra clusters do replicate data across the nodes. The specific number of replicas is configurable, but generally production clusters will use a replication factor of 3. This means that a given row will be stored on three different machines in the cluster (...) In terms of servicing requests, if a node receives a request for data that it does not have it will forward that request to the nodes that do own the data}”.

\item\textit{Explanation of Web server architecture} provides the functionalities or difference between the architecture of Web servers, for example, \textit{Web server functionality explanation} (such as XAMPP, IIS, and WAMP servers) (see Figure \ref{TaxonomyOfArchSolution}). For instance, the first answer in this ARP\footnote{\url{https://tinyurl.com/ysacs8zd}} provides detailed difference between \textit{XAMPP}, \textit{WAMP}, and \textit{IIS} servers as the solution to the question that asks about the difference between those three types of Web severs by expressing that: “\textit{(...) Their (XAMPP and WampServer) differences are in the format/structure of the package, the configurations, and (...) IIS is a web-server application just like Apache is, except it's made by Microsoft and is Windows only (Apache runs on both Windows and Linux) (...)}”.
\end{itemize}

\textbf{(4) Architecture tactic}: This category of ARP solutions provide and explain architecture tactics that enable the realization of specific quality attribute (e.g., performance and security) of software systems. Four subcategories of architecture tactics were identified in this category, among which \textit{tactic for performance} is the most (21 out of 37 ARP solutions) discussed subcategory (see Figure \ref{TaxonomyOfArchSolution}).

\begin{itemize} 
\item\textit{Tactic for performance} provides and explains architecture tactics that assist in the realization of the system performance requirements. We identified four architecture tactics for performance, such as \textit{scheduling resources tactic} and \textit{in memory caching tactic} (see Figure \ref{TaxonomyOfArchSolution}). Regarding \textit{in memory caching tactic}, a developer wanted to choose a suitable design technique between two data handling design techniques (i.e., working directly with a database or working with objects and letting the ORM handle the storage) in order to boost the performance of the inventory system that should handle thousands of item types and quantities of each item stored in a database. According to the scenario elaborated in the question, the first answer in this ARP\footnote{\url{https://tinyurl.com/289ffurv}} suggests to apply \textit{in-memory caching} architecture tactic with ORM to have the system performance boosted by saying that “\textit{(...) most of the time it is easier to do an SQL query, but an in-memory cache can really BOOST performance. Yes, it uses memory. Who cares? Workstations can have 64GB memory these days (...)}”. 

\item\textit{Tactic for maintainability} covers architecture tactics that enable the maintainability requirements of systems. We identified three maintainability tactics, such as \textit{less coupling tactic} and \textit{encapsulation through API introduction tactic} (see Figure \ref{TaxonomyOfArchSolution}). Concerning the \textit{less coupling tactic}, a developer needed to build a scalable, maintainable, and low-latency single sign-on for all web applications. The first answer in this ARP\footnote{\url{https://tinyurl.com/22ckrdv4}} suggests to apply \textit{less coupling tactic} when designing the applications in order to make them maintainable by saying that: “\textit{I would not integrate the authentication on the database level (...) This might become hard to maintain. I would prefer a loosely coupled approach by exposing a simple service on your central server that lets the other app servers run authentication requests (...)}”.

\item\textit{Tactic for security} provides architecture tactics that help the realization of the system security requirements. This subcategory includes three architecture tactics, \textit{authentication tactic}, \textit{limiting access tactic}, and \textit{communication link encryption tactic} (see Figure \ref{TaxonomyOfArchSolution}). Regarding \textit{authentication tactic}, the first answer in this ARP\footnote{\url{https://tinyurl.com/bdhk4uud}} provides and explains \textit{authentication tactic} to a question that asked for how to set up two level authentication approaches of the ‘user JWT’ in microservice based application by stating that: “\textit{(...) You can achieve the two levels of security you require by using a single user token and claims based authorisation. If a call is made to the gateway with the user token, the gateway authenticates the call based on the user token, retrieves the ‘userId’ claim (...)}”. 

\item\textit{Tactic for availability} collects architecture tactics that enable the system availability requirements. We collected three availability tactics, like \textit{data replication tactic} and \textit{functional redundancy tactic} (see Figure \ref{TaxonomyOfArchSolution}). Concerning \textit{data replication tactic}, a developer asked how to achieve the availability requirement for an application that needs to use two Amazon EC2 instances each with Cassandra database. The first answer in this ARP\footnote{\url{https://tinyurl.com/2p8db7mu}} provides and explains the replication mechanism that could be applied in his/her application (according to the design scenario described in the question) by saying that: “\textit{(...) In your scenario (since you are in a single DC) you can use SimpleStrategy for your replication strategy and a Replication Factor (RF) of 2. With this setup, you will have all data replicated on both nodes. This will make the data available from either node with a covet}”.
\end{itemize}

\textbf{(5) Architecture pattern}: The ARP solutions in this category provide architecture patterns for addressing multiple system quality attributes, and also provide commonly used architecture patterns in certain application domains (see Figure \ref{TaxonomyOfArchSolution}). Among the two subcategories identified in this category, the \textit{architecture pattern suggestion to meet quality attribute} subcategory contains the majority of solutions (17 out of 30 ARP solutions of the \textit{architecture pattern} category) that SO users discuss in this category (see Figure \ref{TaxonomyOfArchSolution}). 

\begin{itemize} 
\item\textit{Architecture pattern suggestion to meet quality attributes} collects architecture patterns for addressing system quality attributes, such as \textit{patterns for modifiability, reusability, and portability} (Broker, MVC, and SOA) (see Figure \ref{TaxonomyOfArchSolution}). For example, a SO user asked about the best C\# architecture patterns enabling the communication between separate plugins of a multi-tenant website wherein modifiability, reusability, and flexibility are the major concerns. The first answer in this ARP\footnote{\url{https://tinyurl.com/42m8ts56}} recommends \textit{SOA pattern} as a solution to that question by noting that: “\textit{(...) I might suggest Service Oriented Architecture. Mostly because it can bend to a business in a very quick and agile manner. This architecture provides many bonuses: Lightweight, Agile, Code Re-usability (...)}”.

\item\textit{Usage of architecture pattern} gathers architecture patterns for questions that ask about the commonly used architecture patterns in certain application domains (see Figure \ref{TaxonomyOfArchSolution}), such as \textit{pattern for time-critical system} (Preemptive Multitasking), \textit{pattern for Android application} (MVP, MVVM, MVC, Observer), and \textit{pattern for distributed system} (SOA, Broker). For example, a user asked about architecture pattern for time-critical applications. The first answer in this ARP\footnote{\url{https://tinyurl.com/25y4b9e6}} recommends \textit{Preemptive Multitasking} pattern to that question by saying that: “(...) \textit{This pattern is called preemptive RTOS, which is capable of handling the events immediately (...)}”.
\end{itemize}

\textbf{(6) Architecture solution for deployment} collects the ARP solutions that discuss the deployment of architecture of systems in the hosting devices (either on the Cloud or the local server) in order to address the systems' requirements. This category consists of two subcategories, among which \textit{process for deployment} is the most (9 out of 13 ARP solutions) discussed subcategory (see Figure \ref{TaxonomyOfArchSolution}).

\begin{itemize}
\item\textit{Process for architecture deployment} collects the ARP solutions that discuss the processes for deploying the architecture of applications for the purpose of achieving the applications' requirements (e.g., functional or nun-functional requirements). We identified several processes for architecture deployment, for example, \textit{deployment process with Azure App service}, \textit{deployment process with Kubernetes}, and \textit{deployment process with AppHabor for .NET applications} (see Figure \ref{TaxonomyOfArchSolution}). Regarding \textit{deployment process with Kubernetes}, a responder provided and explained the deployment process with Kubernetes to an asker who wanted to deploy a microservices architecture (which was built up with 15 Spring Boot microservices) on five Kubernetes nodes with one cluster master. According to the scenario described in the question, the first answer in this ARP\footnote{\url{https://tinyurl.com/483en2ts}} suggested to use three cluster masters at a minimum instead of one cluster master in order to avoid the data loss and consequently address the system's availability requirement by saying that: “\textit{(...) one master is not enough. The loss of that VM, the underlying hardware, or a failure of the services on the master will lead to an outage for all customers and potentially catastrophic data loss. Run 3 masters at minimum}”.

\item\textit{Tool for architecture deployment} collects the tools for deploying architecture of systems in order to achieve the requirements of the systems. We collected several tools, such as \textit{simulator tool for architecture deployment} and \textit{tool for continuous deployment} (see Figure \ref{TaxonomyOfArchSolution}). Regarding the \textit{tool for continuous deployment}, the first answer in this ARP\footnote{\url{https://tinyurl.com/yc7j8z5j}} recommends Argo CD tool as the solution to the question that asks about a tool for microservices architecture continuous deployment on Kubernetes by stating that: “\textit{(...) ArgoCD workflow provides that functionality (...)}”.
\end{itemize}

\textbf{(7) Solution for architecture documentation}: The ARP solutions in this category provide the approaches and tools that enable the documentation of architecture (see Figure \ref{TaxonomyOfArchSolution}). This category consists of two subcategories, among which \textit{tool for architecture documentation} is the most (3 out 5 ARP solutions) discussed subcategory (see Figure \ref{TaxonomyOfArchSolution}).

 \begin{itemize}
 \item\textit{Tool for architecture documentation} suggests the tools that can facilitate the documentation of architecture, such as \textit{tool for Web application architecture documentation} and \textit{tool for database architecture documentation} (see Figure \ref{TaxonomyOfArchSolution}). Concerning the \textit{tool for Web application architecture documentation}, the first answer in this ARP\footnote{\url{https://tinyurl.com/4zm82snw}} suggests NJsonSchema tool as the solution to the question that asks about a tool for documenting a microservices-based application by saying that: “\textit{(...) there is NJsonSchema tool https://github.com/NJsonSchema/NJsonSchema}”. 
 
\item\textit{Viewpoint for architecture documentation} provides the viewpoints for architecture documentation, such as \textit{development viewpoint for architecture documentation}, and \textit{deployment viewpoint for architecture documentation}. For example, the first answer in this ARP\footnote{\url{https://tinyurl.com/2p8ezw7j}} provides two viewpoints for architecture documentation (i.e., \textit{development viewpoint for architecture documentation} and \textit{deployment viewpoint for architecture documentation}) for documenting an architecture that is implemented with Java.
\end{itemize}

\begin{tcolorbox}[colback=gray!5!white,colframe=gray!75!black,title=Key Findings of RQ5]
\textbf{Finding 7}:
SO users frequently use two terms related to usefulness (i.e., “useful” and “helpful”), along with six other terms (i.e., “definitely”, “very”, “really”, “super”, “extremely”, and “incredibly”) in the comment threads to explicitly express their feedback about how useful they found certain architecture solutions provided to their associated architecture related questions.

\textbf{Finding 8}:
We derived a taxonomy of useful architecture solutions consisting of 7 categories, 20 subcategories, and 85 types, indicating the diversity of useful architecture solutions provided in SO.

\textbf{Finding 9}:
\textit{Solution for architecture configuration} (39\%, 127 out 324 ARP solutions), \textit{solution for architecture implementation} (18\%, 59 out 324 ARP solutions), \textit{explanation of architecture} (16\%, 53 out 324 ARP solutions), and \textit{architecture tactic} (11\%, 37 out 324 ARP solutions) are the top four most frequently discussed categories of useful architecture solutions. 
 \end{tcolorbox}

\subsection{Characteristics of useful architecture solutions (RQ6)} \label{resultsofRQ6}
\begin{figure}[h]
 \centering
 \includegraphics[width=0.7 \linewidth]{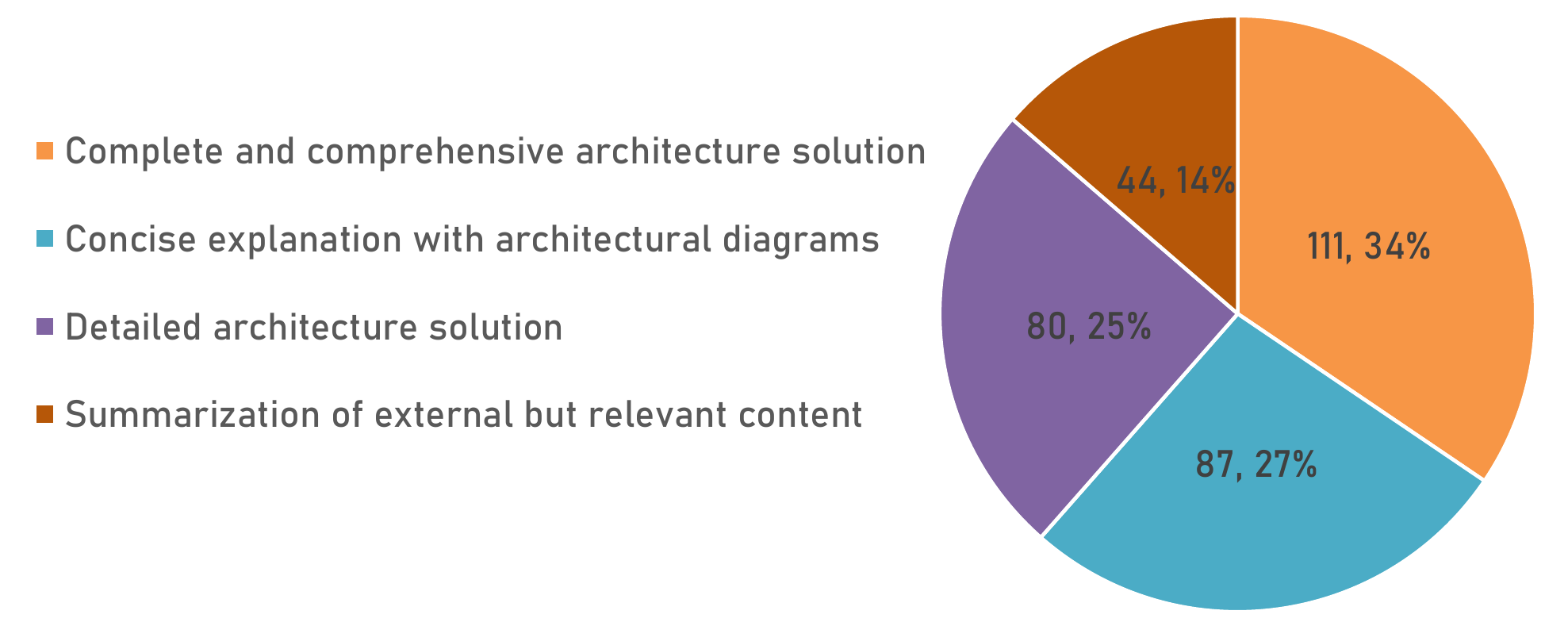}
 \caption{The common identified characteristics of architecture solutions that are considered useful}
 \label{CharacteristicsOfusefulSoluions}
\end{figure}

As shown in Figure \ref{UsefulComment}, SO users occasionally leave comments under an architecture solution to convey that such solution is useful. Hence, this motivated us to study the characteristics of the architecture solutions that are considered to be useful. Analogous to RQ5, we used the 324 ARPs (a subset of 968 ARPs) with useful knowledge (see Figure \ref{DrawingStudyExcution}) to analyze the architecture solutions and their attached comments, and study the characteristics of those solutions. The qualitative data analysis (see Section \ref{dataAnalyis}) identified four common characteristics of architecture solutions that are considered useful by SO users. Figure \ref{CharacteristicsOfusefulSoluions} depicts these characteristics along with their counts, in which \textit{complete and comprehensive architecture solution} appears to be the most (34\%, 111 out of 324 ARP solutions) frequent characteristic of architecture solutions that SO users consider to be useful. 

\textbf{(1) Complete and comprehensive architecture solution}: A developer may ask more than one question (sub-questions) in one single architecture related question in SO. A solution that addresses all sub-questions asked in the question and provides comprehensive responses (e.g., providing rationale, such as benefits and drawbacks of the provided architecture solution) to these sub-questions is considered useful. For example, a developer posted this comment: “\textit{+1 for the most complete, comprehensive useful response I've ever seen (...)}” under the first answer in this ARP\footnote{\url{https://tinyurl.com/dfw27h38}} that comprehensively addresses all sub-questions (e.g., difficult to visualize data in the system architecture implemented with Java and Python) asked in the question. 

\textbf{(2) Concise explanation with architectural diagrams} provides a brief explanation about the key elements of the architecture solution. Some examples of these key elements could be the best architecture patterns, tactics, and technologies (e.g., databases) to be used in order to address the design concerns described in the question. In addition, providing architectural diagrams, such as component diagrams to represent and summarize the practical applicability of the solution also contributes to the architecture solution being considered useful. For example, one developer asked whether “command handler” and “command bus” should belong to or be implemented in the application layer or domain layer in the architecture. At first, in the first answer of this ARP\footnote{\url{https://tinyurl.com/yh292pn8}}, a responder provided a concise and relevant solution. But the asker was not satisfied with this solution and then s/he commented to request a sequence diagram (which was provided later) to be associated with the solution for it to be useful: “\textit{Thanks, David. It would be really useful if you could share a sequence diagram. Appreciate it}”.

\textbf{(3) Detailed architecture solution}: These solutions provide and fully describe all necessary architectural elements (such as patterns, components) and other various aspects to be considered (e.g., solution trade-offs, constraints, and alternatives) when addressing the design concerns stated in the question. For instance, a developer posted this comment: “\textit{Thank you for your detailed answer. This is certainly very helpful}” under the second answer in this ARP\footnote{\url{https://tinyurl.com/pfezm7nn}} that lists and details all necessary architectural elements (e.g., quality attributes) and other aspects (e.g., pros and cons of the solution, and alternative solutions) that should be considered when addressing the design concerns (e.g., integrating external modules (external Web applications) into Drupal or vice versa) stated in the question. 

\textbf{(4) Summarization of external but relevant content}: Answer seekers do not like to have external links (URLs) only as solutions posted to their questions since the links may die and the solutions become not accessible and useless. During our data analysis, we observed that answer seekers prefer to have the relevant content summary of URLs instead of the URLs only for the architecture solutions. For example, the first answer in this ARP\footnote{\url{https://tinyurl.com/2p87vu99}} summarizes the content from three URLs to answer the question which mainly asks about the design approach to follow in order to address system availability with Cassandra database. A developer commented under the answer: “\textit{Thank you very much for your information. Your Explanation is sufficient and the links you mentioned are very useful}”.

\begin{tcolorbox}[colback=gray!5!white, colframe=gray!75!black,title=Key Findings of RQ6]

\textbf{Finding 10}:
\textit{Complete and comprehensive architecture solution} is the most (34\%, 111 out of 324 ARP solutions) frequent characteristic of architecture solutions that SO users consider to be useful.

\textbf{Finding 11}:
The presence of architectural diagrams (e.g., components diagrams) in the provided architecture solutions increases the chance of these solutions to be considered useful.
 \end{tcolorbox}

\section{Discussion}\label{discussionAndImplications}

In this section, we revisit the findings of this study by interpreting the results in Section \ref{analysisoftheresults} and discussing their implications for various stakeholders in Section \ref{implications}. 

\subsection{Analysis of the results}\label{analysisoftheresults}

\subsubsection{The delta between our results and the results from prior work}

Similar to our study, several studies have analyzed ARPs from SO to mine architectural knowledge discussed by SO users in order to support architecting activities. In this section, we discuss the relationship and difference between our study results and the results in the prior studies (i.e., the three studies by Soliman \textit{et al.} \cite{soliman2016architectural}\cite{soliman2017developing}\cite{soliman2018improving}), which are closely related to our work. 

Soliman \textit{et al.} \cite{soliman2016architectural} identified and analyzed ARPs from SO that discuss architecture knowledge with a focus on technology decisions (one type of architecture decision~\cite{kruchten2004ontology}). They classified these ARPs based on two dimensions: \textit{the purpose of the question} and \textit{the solution type of the question}. They further classified the \textit{purpose dimension} into three subtypes: solution synthesis, solution evaluation, and multi-purposes, and the \textit{solution type dimension} into three subtypes: technology feature, technology bundle, and architecture configuration. In total, their analysis generated 6 types of ARPs. Our analysis generated 9 categories and 21 subcategories of ARP questions (see the results of RQ1 in Table \ref{CategoriesOfQuestions}), such as architecture configuration, architecture decision, architecture concept, architectural implementation, architecture evolution, and architecture refactoring. Some of the types of ARPs (e.g., solution synthesis, solution evaluation, architecture configuration) found by Soliman \textit{et al.} in \cite{soliman2016architectural} are aligned with some of our ARP types, and most of the types of ARPs presented in \cite{soliman2016architectural} can be subcategories of the main categories reported in our work. For example, we have a main category encoded \textit{architecture decision}, and this category can cover three types of APRs (solution synthesis, solution evaluation, and multi-purposes) reported in \cite{soliman2016architectural}. Moreover, our analysis generated new categories of ARPs (such as architecture concept, architecture tool, architecture evolution, architecture refactoring, architecture deployment, and architecture documentation).

Soliman \textit{et al.} \cite{soliman2017developing} used the same sample of ARPs that were used in their previous work (i.e., \cite{soliman2016architectural}) and developed an ontology that covers architectural knowledge concepts in SO. The ontology consists of three main ontology classes: \textit{simple ontology class}, \textit{composite ontology class}, and \textit{lexical trigger ontology class}. A simple ontology class is composed of subclasses, for example, technology solution, architecture pattern, quality attribute, architecture component, and architecture connector. The composite ontology class consists of several subclasses, such as architecture configuration, technology feature, technology benefits and drawbacks, technology user-case, user request, and design rule. The lexical trigger ontology class has subclasses, such as difficulty adjectives, advise verbs, value adjectives, wish verbs, support verbs, versus prepositions. Some subclasses found by the analysis in \cite{soliman2017developing}, such as architecture configuration and architecture pattern, are aligned with our results of RQ5 (see Figure \ref{TaxonomyOfArchSolution}). However, the analysis in \cite{soliman2017developing} is based on a sample of ARPs that mainly discuss technology information (e.g., requirements and constraints on technology solutions, technology benefits and drawbacks, and technology features). Our analysis complements the work in \cite{soliman2017developing} by adding several new categories, such as architecture tactic, explanation of architecture, solution for architecture documentation, and solution for architecture deployment, leading to more comprehensive categories and subcategories of ARP solutions provided in SO.

Soliman \textit{et al.} \cite{soliman2018improving} developed a search approach that relies on the classification approach to provide suitable types of ARPs for each design step proposed by Kazman and Cervantes \cite{cervantes2016designing}. The analysis conducted by Soliman \textit{et al.} \cite{soliman2018improving} is also based on the sample of ARPs from their previous work \cite{soliman2016architectural}, and some other posts extracted from SO. Specifically, their search approach classifies SO posts into four types: technology identification, technology evaluation, features and configuration, and programming posts. The first three types of posts are ARP types that were reported in their previous study (i.e., \cite{soliman2016architectural}), and in the first paragraph of this section, we have already described the difference and similarities between these types of ARPs in \cite{soliman2016architectural} and the types of ARPs in our study (see results of RQ1 in Table \ref{CategoriesOfQuestions}). 

In addition to the abovementioned difference between our study results and the results reported in \cite{soliman2016architectural}\cite{soliman2017developing}\cite{soliman2018improving}, in our study, we investigated a new set of research questions (RQ2, RQ3, RQ4, RQ5, and RQ6). We explored other types of architecture knowledge, such as design contexts (RQ2) discussed in architecture related questions, characteristics of ARPs (questions and solutions) (RQ3, RQ4, and RQ6), and the usefulness of the ARP solutions (RQ5), which was not the concern of the abovementioned studies (i.e., \cite{soliman2016architectural}\cite{soliman2017developing}\cite{soliman2018improving}). Moreover, our analysis covered the entire post, including the question and its associated comments (RQ3, RQ4), the answers to the question and their associated comments (RQ5, RQ6). The analysis in the abovementioned studies by Soliman \textit{et al.} only focused on questions and answers. Thus, our study results add new information to the state of the art, and practitioners and researchers can benefit from our study results and findings (e.g., the taxonomy of architecture solutions that are considered useful).

\subsubsection{Identified categories of ARPs in SO could support architecting activities}

The significant results of this study are categories of ARPs (questions (i.e., RQ1) and solutions (i.e., RQ5)). This study reveals that SO users ask a broad range (nine categories) of architecture related questions, such as questions about \textit{architecture configuration}, \textit{architecture decision}, \textit{architecture concept}, \textit{architecture implementation}, and \textit{architecture tool} (see Table \ref{CategoriesOfQuestions} in Section \ref{resultsofRQ1}). In addition, we classified the architecture solutions that are considered useful into seven categories, such as \textit{solution for architecture configuration}, \textit{solution for architecture implementation}, \textit{explanation of architecture}, and \textit{architecture tactic} (see Figure \ref{TaxonomyOfArchSolution} in Section \ref{resultsofRQ5}). One observation is that our identified categories of these ARPs (questions and solutions) cover almost all the architecting activities that span from the initial stages (e.g., architectural analysis, synthesis, and evaluation \cite{HofmeisterGenModel2007}) of architecture creation to the later stages (e.g., architectural implementation, and maintenance and evolution \cite{tang2010comparative}) in a system lifecycle. Thus the identified categories of architecture related questions and solutions can support the mentioned architecting activities during the architecture lifecycle. These results also support the findings by Soliman \textit{et al}. in \cite{soliman2016architectural} that SO should be considered as one of the important sources of architectural knowledge. Moreover, practitioners reported Q\&A sites (e.g., Stack Overflow) as the most useful when searching architectural information according to our recent industrial survey \cite{musenga2022}. Thus, practitioners could rely well on SO to identify, such as, the benefits and drawbacks of architecture solutions in certain application domains, for example, the benefits and drawbacks about the \textit{framework for iOS applications} in our taxonomy (see Figure \ref{TaxonomyOfArchSolution}) for architecture implementation.

\subsubsection{Importance of design context in architecture design}

The results of RQ2 reveal that in most (71\%, 687 out of 968) of the studied ARP questions, SO users considered the design contexts (i.e., knowledge about the environments in which the systems are expected to operate \cite{bedjeti2017modeling}) when describing the design concerns in their architecture related questions (\textbf{Finding 2} in Section \ref{resultsofRQ2}). One reason could be that SO users prefer to provide a brief description of their project backgrounds and then expect responders to suggest potential architecture solutions with their rationale based on the given design concerns and design contexts. Moreover, the results of RQ2 show that most of the SO users do consider design context as one of the indispensable ingredients that can drive the architecture design of a system \cite{bedjeti2017modeling}. 

\subsubsection{Identified characteristics of ARPs to improve their quality}

From Table \ref{CharacteristicMoreAnswers}, Table \ref{CharactertiscsFewAnswes}, and Figure \ref{CharacteristicsOfusefulSoluions}, we found that there are various characteristics of ARPs (questions and answers) in SO. For example, we observed that architecture related questions that articulate well architectural information are likely to get more than one answer (see Table \ref{CharacteristicMoreAnswers}), while architecture related questions that lack certain significant information and poorly structured (see Table \ref{CharactertiscsFewAnswes}) tend to only get one answer. The reason is the following: well-articulated architecture questions provide an overview of the planned system and describe well their design concerns, which helps potential responders to fully understand the purposes of these questions so that they can provide answers. We also found that answer seekers highly appreciated architecture solutions that are complete and comprehensive and considered them to be useful. One reason is that these architecture solutions address design concerns raised in all sub-questions (in the case that there are sub-questions in one question) by providing comprehensive solutions, for example, design contexts, pros and cons of the provided architecture solutions, which helps the answer seekers understand why such architecture solutions are the way they are. The identified characteristics of ARPs (questions and answers) in SO show that SO users have varying needs in the description of ARPs (questions and answers) and the level of details. These findings could assist in improving the quality of the posted architecture related questions and answers at SO.

\subsection{Implications}\label{implications}

\subsubsection{For Stack Overflow}

\textbf{Increase the awareness of SO towards its users}: SO introduces itself as a community Q\&A platform for asking and collecting programming related knowledge during software development. Thus, the majority of the SO users use the platform as the place for sharing and learning coding related knowledge only. However, ever since this site started growing and being popular, architects have begun to share their competencies, experience, and architecture problems by asking architecture related questions or providing architecture solutions, such as architecture patterns \cite{bi2018architecture}. Akin to searching and (re)using existing code examples provided in SO to solve programming related problems, SO users also search and (re)use existing architecture solutions, such as architecture tactics \cite{bi2021mat} in SO for solving their architecture design concerns (e.g., architecture design to meet quality attributes). Hence, SO not only curates programming related knowledge, but also accumulates architecture solutions provided to a wide range of architecture problems or design problems \cite{soliman2016architectural, bi2021mat}. However, during our study, we found that architecture related questions were being seen as off-topics in SO and should not be asked at the site (see Table \ref{CharactertiscsFewAnswes} in Section \ref{resultsofRQ4}) due to SO users' perception or awareness of what SO is used for (i.e., a site for programming related issues). Given this situation, there is a possibility that interesting architecture related questions asked might remain unanswered or even be deleted by the site moderators. Although some SO users see architecture related questions as off-topic, we think that architecture related questions will sustain and continue to thrive in SO. According to our studied dataset (318 architecture related questions) relevant for answering RQ4 (characteristics of architecture related questions that only have one answer), we observed that architecture related questions that were commented to be off-topic are not many (16\%, 51 out 318 architecture related questions). This finding is promising for the long-term prospect of architecture related questions in SO. Moreover, we argue that architecture related questions which communicate architectural knowledge \cite{10yearsSAKM} are an important type of questions and have a system-wide impact on software development. Many architecture related questions arise during development when addressing specific design concerns (e.g., quality attributes) and their trade-offs. Therefore, architecture related discussions (e.g., through architecture related questions) should not be seen as off-topic in SO, and SO should consider increasing its awareness (i.e., to be a site for development related issues instead of a site for programming related issues only) towards its users and welcome architecture questions to be discussed on the site.

\textbf{Adjust the current answers and comments organization mechanisms to improve the search and (re)use of architecture solutions}: SO attracts a large number of users with different backgrounds, skills, expertise, and viewpoints. Thus during our data analysis, we have observed that an architecture related question like any other questions (e.g., programming related question) in SO may receive multiple (or alternative) answers. The study by Wang \textit{et al}. \cite{wang2018users} reported that nearly 6.5 million questions (37\% of all questions at SO) had more than one answer, and the average length of an answer is 789 characters. With the current SO answers organization mechanism, when there are multiple answers to a question in a single post, at most one answer per question can be accepted/marked by the asker to indicate that the answer is the most useful one \cite{treude2011programmers}. This asker should be a registered user with at least 15 reputation on SO \cite{GoodAcode2012}. The registered users without required reputation (i.e., less than 15 reputation) on the site are restricted from accepting or voting (upvoting or downvoting) answers to indicate that such answers are useful \cite{GoodAcode2012}. Consequently, leaving a large number of answers in SO that are not accepted or marked as useful answers yet being useful, just because the users (askers) do not possess the required minimum reputation to do so. During our data analysis, we observed that not all useful architecture solutions are explicitly marked (i.e., accepted as useful) in SO to facilitate the search and (re)use of those solutions (for example, see Figure \ref{UsefulComment}). Also, we found that SO users may use terms related to usefulness, such as “useful” and “helpful”, in the comment threads to explicitly express their feedback about how useful they found certain architecture solutions provided to their associated questions in SO (\textbf{Finding 7} in Section \ref{resultsofRQ5}). Our finding is in line with the findings by Zhang \textit{et al}. in \cite{readingAnswers2019} and \cite{obsolete2019} that comments provide additional information to support the answers, such as improvement of answers \cite{readingAnswers2019} and obsoleted answers \cite{obsolete2019}. Prior studies criticized the comment organization mechanism at SO (e.g., \cite{zhang2021comments}). In order to keep each answer thread compact, SO implements a comment organization mechanism to only show the top 5 comments \cite{zhang2021comments}. Aiming at showing the most informative comments and hiding less informative ones, the mechanism first ranks these comments based on their scores. When multiple comments have the same score, they are then ranked by their creation time \cite{zhang2021comments}. Hidden comments are not indexed by Google\footnote{\url{https://tinyurl.com/2p87yyfr}}. Thus, due to this current comment ranking mechanism, informative comments might be hidden in turn reducing the chances of someone retrieving or voting on them. Regardless of its success and popularity, navigating SO remains a challenge, and it is insufficient how SO directs its users to retrieve informative comments \cite{nadi2020essential}. Comments that state the usefulness of answers (including architecture solutions) are one of the most important informative comments. Thus, we provide SO the following suggestion:  

\begin{itemize}
\item Instead of simply ranking comments by their score then their creation time \cite{zhang2021comments}, the comments organization mechanism needs to introduce a higher priority for more informative comments. SO may consider adjusting its comments organization mechanism by, for instance, developing special analytical techniques (e.g., machine learning approaches) that could filter and rank comments stating, for example, improvement of answers \cite{readingAnswers2019}, usefulness of answers (e.g., useful architecture solutions).

\item SO may also refer to and extend our proposed taxonomy of useful architecture solutions (solutions commented to be useful) to develop an automated tool that could assist the SO users in identifying existing architecture solutions with, for example, useful knowledge. 
\end{itemize}

\subsubsection{For SO users}

Throughout the qualitative analysis of RQ3, RQ4, and RQ5, we identified various characteristics of ARPs (questions and solutions) in SO. Among these characteristics, we found that questions that provide \textit{clear description together with architectural diagrams} increase their likelihood of getting more than one answer (see Table \ref{CharacteristicMoreAnswers}), while \textit{poorly structured architecture questions} (see Table \ref{CharactertiscsFewAnswes}) tend to only get one answer. Also, we found that architecture solutions that provide \textit{concise explanation with architectural diagram} is the second most common characteristic of architecture solutions that are considered useful (see Figure \ref{CharacteristicsOfusefulSoluions}). One observation is that SO users would like to see architectural diagrams, such as components diagrams, in both questions and solutions as these diagrams can benefit both parts. Concerning questions, providing architectural diagrams increases their chance of getting more responses (e.g., more than one answer) (\textbf{Finding 5} in Section \ref{resultsofRQ3}). On the other hand, architectural diagrams in solutions boost their chances of being considered useful (\textbf{Finding 11} in Section \ref{resultsofRQ6}). Therefore, both askers and responders should better provide diagrams in their ARPs (questions and answers). One reason is that architecture is at a high abstraction level, and it would be hard to describe an architecture problem and much harder to explain an architecture solution with text only. Architectural diagrams make architecture to be more understandable \cite{thomas2013con}, and stakeholders can communicate about architectural problems and solutions more easily using architectural diagrams. Moreover, various identified characteristics of ARPs in this study (see Table \ref{CharacteristicMoreAnswers}, Table \ref{CharactertiscsFewAnswes}, and Figure \ref{CharacteristicsOfusefulSoluions}) are indicators that SO users have varying needs in the formation of both architecture related questions and architecture solutions and the level of details. Therefore, there is a need to provide guidelines to SO users to follow when posting their architecture related questions and solutions.

\textbf{For SO askers}: In the following, we provide the guidelines for SO askers to follow when posting their architectural related questions with more likelihood of being answered by other SO users and get more than one answer from SO users:
\begin{itemize}
\item \textit{Include architectural diagrams with clear description in the questions}: We recommend askers to add architectural diagrams (e.g., component diagrams) and specifically clarify the design concerns in their architecture related questions to help other SO users better understand the purposes of their questions.

\item \textit{Write well-articulated architecture questions with descriptive details about the context}: We suggest that askers could describe well architectural information in their questions. This can be done, for example, by providing an overall understanding of the system, as well as detailed information on components in their scope together with interfaces and relationships to other components. Also, we recommend askers to add information about the design contexts, since design contexts are critical for other SO users to correctly understand your architecture related questions. 
\end{itemize}

\textbf{For SO responders}:
As stated throughout this study, we not only analyzed architecture related questions, but also examined the characteristics of architecture solutions that are considered useful by SO users. Thus, in the following, we provide guidelines to SO users to follow when posting their solutions with more likelihood of being considered useful by other SO users: 

\begin{itemize}
\item \textit{Write concise architecture solutions with architectural diagrams}: Responders are recommended to write concise architecture solutions by stating key points only in the solutions and add architectural diagrams (if necessary) that depict and clarify, for example, the architecture implementation view in their posted solutions. 

\item \textit{Include URLs in architecture solutions with sufficient and relevant architectural knowledge}: Answer seekers do not like to have external links (URLs) only as solutions posted to their questions \cite{GoodAnsw2012}. In the case when a responder wants to make the architecture solution short, s/he can provide links to external websites that contain more explanations or complex examples, and his/her solution should be self-contained. In other words, this solution should provide certain important and relevant architectural knowledge which can make it explainable, such as design decisions and their rationale, contexts, assumption, and other factors that together determine why a particular solution is the way it is.

\end{itemize}

\subsubsection{For researchers}
 
\textbf{Towards innovative tools to search and (re)use architectural knowledge in SO}: The results of our study (e.g., categories of architecture related questions in Section \ref{resultsofRQ1} and their useful solutions in Section \ref{resultsofRQ5}) provide insights into the nature of SO users' discussions on architecture design in SO. In addition, the results of this study re-emphasize the conclusion by Soliman \textit{et al}. \cite{soliman2017developing} that SO should be considered as one of the important sources of architectural knowledge. However, SO captures large amounts of information in its posts and this information is mainly represented as unstructured text. Furthermore, the abstract nature of architectural concepts makes it difficult for keyword-based searches to find architecture relevant information, and this might not be easy for SO users to capture and (re)use the architectural knowledge (e.g., benefits, drawbacks, and trade-offs of using specific architecture patterns in certain application domains) from SO. Therefore, researchers can contribute to improving the search and (re)use of the architectural knowledge in SO by focusing on innovative techniques and tools that could efficiently and effectively guide the capturing and usage of this knowledge to support architecting activities (e.g., architectural analysis and synthesis \cite{HofmeisterGenModel2007}). For example, researchers can refer to our proposed taxonomy of useful architecture solutions in SO as a guidance to develop automated approaches and tools that could mine and locate architecture solutions (e.g., \textit{solution for architecture configuration}, the most common category of useful architecture solutions in SO, see Figure \ref{TaxonomyOfArchSolution}) for addressing similar design concerns (e.g., questions that ask about \textit{architecture configuration}, see Table \ref{CategoriesOfQuestions}). This could help SO users to check the questions and solutions that are relevant to their design concerns (e.g., banking system configuration). Furthermore, we observed that \textit{architecture configuration} (27\%), \textit{architecture decision} (19\%), and \textit{architecture concept} (15\%) are the top three categories of most frequently asked architecture related questions (\textbf{Finding 1} in Section \ref{resultsofRQ1}), and researchers may explore the challenges (that are being faced by SO users) related to these most frequently asked categories of architecture related questions.

\textbf{Investigation of design contexts in Q\&A sites to support architecture knowledge management}: From Table \ref{summaryOfDesignContexts} in Section \ref{resultsofRQ2}, we found that SO users discuss about design contexts along with design concerns when asking architecture related questions in SO. Three categories (application, platform, and organization contexts) and eight subcategories (application domain, external service, software, hardware, development schedule, stakeholders, and resources contexts) of design contexts were identified from our studied sample of ARPs (see Table \ref{summaryOfDesignContexts}). Whilst we know that SO users discuss design contexts along with design concerns when asking architecture related questions, there have been very few studies on mining design contexts in the Q\&A community sites, such as SO, to support architecture knowledge management \cite{wijerathna2022mining}, which is an interesting area to be explored in future studies. 

\section{Threats to validity}\label{ThreatValidity}
In this section, we discuss the threats to the validity of the study results by following the guidelines proposed by Wohlin \textit{et al}. \cite{ref35} and how these threats were mitigated in our research.

\textbf{Internal validity} concerns with the selection of search terms used to mine ARPs in SO. We used search terms, such as “architecture” and “architectural”, to identify the related posts in SO (see Section \ref{studyExcution}), and this is a threat to the internal validity in our study because we might have missed other terms, such as “design”, that SO users use to express architecture concepts. Hence, the search terms we used in this study may not be able to identify the complete set of ARPs in SO. To reduce this threat, we first conducted a pilot search and observed that SO users use the term “design” mostly in the programming context (e.g., “singleton design pattern”\footnote{\url{https://tinyurl.com/8yks7nhm}}). In addition, as mentioned in Section \ref{studyExcution}, using the search terms (e.g., “architecture” and “architectural”) to only search exclusively through tags can be ineffective, because tags can be sometimes less informative \cite{barua2014developers} (see the example provided in Section \ref{datacollection}). Thus, we decided to add the titles and bodies of the questions into the search. In this way, we sought to minimize the risk of missing ARPs that use incorrect or irrelevant tags. Finally, we gathered 10,423 ARPs through the search which is quite a large dataset, and it may not be realistic to thoroughly analyze this size of dataset with human effort in order to get accurate and comprehensive results from this dataset. Hence, we computed a statistically representative sample \cite{israel1992determining} of these 10,423 ARPs and randomly selected 968 ARPs as the dataset to be analyzed in this study. However, to further mitigate this threat, we downloaded and utilized the current SO data dump (i.e., Stack Exchange data dump on October 5, 2022\footnote{\url{https://archive.org/details/stackexchange_20221005}}). This data dump is a snapshot of the underlying database used by SO and it stores all the information for the questions, answers, tags, comments, votes, and user histories in XML files (e.g., \texttt{Posts.xml}). We used \texttt{Posts.xml} file, which stores the questions and answers of all the SO posts, as the basic to estimate how many ARPs we missed due to limiting the search to the ``architect*'' terms in our study. According to the SO data dump of October 5, 2022, there are 23 million questions (posts) and 34 million answers. We then used the power statistics and calculated a representative sample size of these 23 million posts. With a 95\% confidence level and 3\% margin of error, the representative sample size calculated is 1069 posts. Afterwards, we randomly selected 1069 posts from the 23 million posts and manually checked them for calculating how many ARPs we might have missed due to limiting the search to the “architect*” terms during the search of ARPs. Specifically, the first author labelled the 1069 posts to determine which of the posts are ARPs. The second author checked and validated the labeling results. The disagreements were resolved in a meeting to improve the reliability of the labeling results. Based on our manual labelling, we found that out of the 1069 posts, only 21 were ARPs (i.e., the true positives), wherein 14.3\% (i.e., 3 out of 21 ARPs) do not contain “architect*” terms and 85.7\% (i.e., 18 out of 21 APRs) contain “architect*” terms. Therefore, we admit that we might have missed certain number of ARPs (i.e., 14.3\%) that do not contain “architect*” terms. We added in our replication package \cite{dataset} the randomly selected posts (i.e., 1069 posts) and the labeling results (i.e., 18 ARPs which contain “architect*” terms and 3 ARPs which do not contain “architect*” terms) for replication purpose.

\textbf{Construct validity} refers to the degree to which a study measures what it claims to be measuring \cite{ref35}. One threat to the construct validity in this study is concerning the manual analysis of the selected SO posts. This is because manually analysis could bring personal bias due to multiple interpretations and/or oversight. To mitigate this threat, we used two qualitative techniques (open coding and constant comparison) from Grounded Theory \cite{stol2016grounded} to analyze the extracted data and answer the RQs. Moreover, we tried to minimize this threat by performing a pilot data coding before the formal data coding. As discussed in Section \ref{dataExtractionAndAnaly}, during the pilot data coding, the first author selected a random set of 100 ARPs and encoded the extracted data (see Table \ref{dataExtraction}) with respect to the purpose of each RQ (see Table \ref{researchQuestions}). Several physical meetings with the second author were scheduled to solve any confusion faced by the first author during this pilot data coding. Moreover, the final results (i.e., concepts, categories, and subcategories) from the pilot data analysis were checked and validated by other three authors (the second, third, and fourth authors) of this study. The disagreements were resolved in a meeting using the negotiated agreement approach~\cite{campbell2013coding} to improve the reliability of the pilot data analysis results. Another threat is related to the identification of ARPs (solutions) with useful knowledge by checking the comments attached to these posts in order to answer RQ5 and RQ6 (see Phase II, in Section \ref{researchdesign}). To mitigate this threat, as we explained in Section \ref{researchdesign}, we did not count on the occurrence of the terms, e.g., “useful” (and the similar) stated in comments to measure the usefulness of an architecture solution given to certain architecture related question in our study. We rather referred to the usefulness related information in the comments attached to the solutions to investigate SO users' discussions on the usefulness of architecture solutions. In other words, we judged the usefulness using the reaction of the SO users after seeing and using the architecture solutions (see a comment in Figure \ref{UsefulComment}). In addition, we (four authors of this study) first read the solutions (from our studied representative sample) commented to be useful to see whether there are really useful to address the questions \cite{zhu2009multi} before we decided to include such posts (solutions) with useful knowledge for analysis. Thus, we believe that we have adequately mitigated this threat. 

\textbf{External validity} refers to the extent to which the findings of the study can be generalized in other settings \cite{ref35}. In this research, we only used SO as the source to investigate ARPs and their usefulness. Even though SO is a widely used and popular developer Q\&A site, this unique source still poses a threat to the diversity of the study results. To mitigate this threat, our research could be further enhanced by including more sources (e.g., GitHub) and look at architecture related questions to understand the architecture design issues that are being faced by architects and developers. Also, researchers might consider going to the fields and asking for feedback directly from architects and developers to better understand the problems they are facing about architecture design and what architecture solutions can be regarded as useful.
  
\textbf{Reliability} refers to whether the study will provide the same results and findings when it is replicated by other researchers \cite{ref35}. In this study, this threat is largely related to the process of manual data collection and analysis. To mitigate this threat, we (the authors of this study) followed a rigorous procedure that is consisted of data collection and analysis activities (see Section \ref{studyExcution}). Moreover, the results from the classification and characterization stages were cross-checked by involving the four authors of the study. To guarantee the reliability of our results and findings, a replication package, containing the dataset used and the encoded data produced in this work, has been made available \cite{dataset}, allowing other researchers to evaluate the rigor of the design and replicate the study. With these measures, this threat has been partially reduced.

\section{Related work}\label{relatedwork}
The research related to our work comes from studies that investigate software development knowledge in Q\&A sites, such as SO. In this section, we summarize relevant work in two categories: (1) investigation of architectural knowledge in Q\&A sites and (2) quality assessment of the knowledge in Q\&A sites.

\subsection{Architectural knowledge in Q\&A Sites}
A few number of existing studies have studied architectural information provided in ARPs in SO from different perspectives. Bi \textit{et al}. \cite{bi2021mat} used a semi-automatic dictionary-based mining approach to extract Quality Attribute (QA) and Architecture Tactic (AT) related discussions in SO posts. Specifically, they applied the dictionary-based classifier Support Vector Machine (SVM) to automatically identify QA-AT related discussions from SO posts. Moreover, the authors went on to manually structure the design relationships between Architectural Tactics (ATs) and Quality Attributes (QAs) used in practice and build a knowledge base of how developers use ATs with respect to QA concerns from related discussions. Such knowledge can help architects better make ATs design decisions. Chinnappan \textit{et al}. \cite{chinnappan2021architectural} extracted data from five open sources of software repositories, including Stack Overflow and qualitatively mined architectural tactics for energy-efficiency robotics software applied by practitioners in real robotics projects. To foster the applicability of the identified tactics (even beyond the robotics software community), they describe them in a generic, implementation independent manner by means of diagrams inspired by the UML component and sequence diagram notation. The presented energy-aware tactics can serve as guidance for roboticists, as well as other developers interested in architecting and implementing energy-aware software. Soliman \textit{et al}. \cite{soliman2021exploring} conducted an empirical study with 50 software engineers, who used Google to make design decisions using the Attribute Driven Design steps \cite{cervantes2016designing}. Based on the relevance and Architecture Knowledge (AK) concepts specified by software engineers, they determined how effective web search engines are to find relevant architectural information from various sources (including Stack Overflow) and to capture AK. In another work, Soliman \textit{et al}. \cite{soliman2017developing} developed an ontology that covers AK concepts in SO. The ontology provides a description of architecture related information to represent and structure AK in SO. 

Soliman \textit{et al}. \cite{soliman2018improving} also leveraged SO with the goal of improving how architects search for architecturally relevant information in online developer communities. They developed a new search approach (i.e., a domain specific-search approach) for searching architecturally relevant information using SO. They found that the new search approach outperforms the conventional keyword-based search approach (searching through the search engines, such as Google). Tian \textit{et al}. \cite{tian2019developers} conducted an empirical study of SO users’ conception of architectural smells by analyzing the discussions from architecture smell related posts in SO. They found that SO users often describe architectural smells with some general terms, such as “bad”, “wrong”, “brittle” or violation of architecture patterns. Li \textit{et al}.\cite{li2021understanding} extracted data from eight most popular online developer communities, including Stack Overflow, to investigate how developers perceive the notion of architecture erosion, its causes and consequences, as well as tools and practices to identify and control architecture erosion. Among other major findings reported in their study, Li \textit{et al}. found that developers either focus on the structural manifestation of architecture erosion or on its effect on run-time qualities, maintenance and evolution; alongside technical factors, architecture erosion is caused to a large extent by non-technical factors. Zou \textit{et al}. \cite{zou2017towards} used the topic model technique to study non-functional requirements related to textual content in SO posts in order to understand the actual requirements of developers. Our study complements the abovementioned work since it focuses on the investigation of architectural knowledge in SO through the characterization and categorization of architecture related posts. In addition, we examine the usefulness (i.e., are the answers useful to address the questions? \cite{zhu2009multi}) of these posts from the point of view of SO users.

The work closely related to ours is the study by Soliman \textit{et al}. \cite{soliman2016architectural}, which leveraged SO to categorize ARPs based on technology related information provided in those posts. The main difference between our study and their work is the fact that we look at the problems from a wider scope. In other words, our study aims to categorize ARPs in SO by looking at various architectural information, such as architecture tactics, provided in those posts, rather than limiting ourselves to one particular information (i.e., technology information). Therefore, our work complements the work in~\cite{soliman2016architectural} by digging deeper into architecture related posts, for example, identifying additional categories of ARPs and exploring design contexts of architecture related questions. Moreover, we characterize and analyze the usefulness of these posts for practitioners and researchers.

\subsection{Quality assessment of knowledge in Q\&A Sites}
The Q\&A platforms, such as SO, play a significant role in knowledge sharing; however, they still face significant challenges to ensure the quality of their knowledge. This is evident in the growing number of studies that focus on analyzing the quality of the content in the programming related posts in SO from different views, such as code and text.

Dagenais and Robillard \cite{dagenais2012recovering} conducted an empirical study on the traceability links between an API and its learning resources in SO. They found that the majority of API names (89\%) in code snippets from online forums are vague and cannot be easily solved due to the deficiency of code snippets. An \textit{et al}. \cite{an2017stack} studied 399 Android applications and revealed 1,279 cases of copyright violations of code reuse between GitHub and SO. Fischer \textit{et al}. \cite{fischer2017stack} assessed the security-related matters of code snippets in SO and discovered that 29\% are insecure. Zhang \textit{et al}. \cite{obsolete2019} investigated obsolescence of answers in SO and found that 31\% may have potential API usage violations that could yield unexpected behavior, such as system crashes and resource leaks. Ragkhitwetsagul \textit{et al}. \cite{ragkhitwetsagul2019toxic} investigated Java code snippets in SO and identified that 153 clones were copied to SO wherein 66\% were obsolete. Zagalsky \textit{et al}. \cite{zagalsky2012example} presented Example-Overflow, a code search and recommendation tool to suggest high-quality code by using the knowledge in SO. Zhang \textit{et al}. \cite{zhang2018code} conducted an exploratory study on the prevalence and severity of API misuse in SO. Treude and Robillard \cite{treude2017understanding} surveyed GitHub users to comprehensively study the difficulty of code snippets in SO. They found that less than half of the SO snippets are self-explanatory. Ragkhitwetsagul \textit{et al}. \cite{agkhitwetsagul2018awareness} conducted an online survey to investigate the answer obsolescence matter in SO. Their survey results indicated that half of the top answerers are aware of obsolete code examples. However, users rarely and even never fix obsolete code examples. Treude and Robillard \cite{treude2016augmenting} developed a tool to improve API documentation with the use of “insight sentences” extracted from SO. Wong \textit{et al}. \cite{wong2013autocomment} proposed an AutoComment tool to automatically generate comments for Java and Android tagged Q\&A posts in Q\&A sites. The results indicate that accurate, adequate, concise, and useful generated comments help users understand the code. Gao \textit{et al}. \cite{gao2015fixing} investigated questions (with similar crash traces) to automatically fix recurring crash bugs in Q\&A sites. McDonnell \textit{et al}. \cite{mcdonnell2013empirical} investigated APIs evolution in Android ecosystem using the version history data found in GitHub. Their results revealed that Android is evolving fast at a rate of 115 API updates per month on average. Dalip \textit{et al}. \cite{dalip2013exploiting} suggested a method to rank answers with regard to the feedback provided to answers. They witnessed that both user and review features are essential to assess the quality of answers. Xu \textit{et al}. \cite{xu2017answerbot} proposed an approach called AnswerBot to automatically summarize answer posts relevant to a technical question in SO. Zhu \textit{et al}. \cite{zhu2009multi} proposed a multi-dimensional model for assessing the quality of answers in social Q\&A sites, such as Answerbag and Yahoo! Answers, in the context of eLearning. Calefato \textit{et al}. \cite{calefato2019empirical} conducted an empirical study aimed at assessing 26 best-answer prediction models in SO.

These studies are related to our work since they investigated the quality of code examples provided in SO posts, while our work investigates the quality (i.e., usefulness) of the architecture solutions provided in SO posts. Nevertheless, our work differs from the aforementioned work in that they focused on low-level source code (e.g., API), while our study focuses on high-level concepts (e.g., proposed architecture patterns as solutions to address design concerns) to investigate their usefulness. We believe that our study complements the existing work on the quality of SO posts by analyzing architecture related posts.

To the best of our knowledge, there has been no investigation of the architectural information provided in ARPs with regard to their categories, characteristics, and usefulness (i.e., are the answers useful to address the questions?) from the point of view of SO users, which is the focus of this study.

\section{Conclusions and future work} \label{conclusionFurtureWork}

Investigating architecture solutions (e.g., architecture tactics and patterns) as an important type of architectural knowledge provided in online developer communities, such as SO, is crucial since this knowledge is one of the most important development knowledge \cite{SA2012}. Architectural knowledge plays a significant role for architects and developers in making informed architectural design decisions during development \cite{jansen2005SoftArch}. Architecture solutions are the fundamental building blocks in modern software design \cite{SA2012}. Contrarily to changing implementation (e.g., low-level source code), once an architecture solution (e.g., an architecture pattern) is adopted and implemented, it is quite difficult and costly to change it \cite{SA2012}. By analyzing and understanding how SO users deal with architectural problems or issues in online developer communities, such as SO, brings three benefits: (1) it provides key insights about the types of design problems SO users face during their architecture designs and the types of architecture solutions discussed as well as their usefulness, (2) it can help to know the design contexts in which architecture problems are raised, and (3) it can help to know the characteristics of architecture problems and solutions discussed. These benefits provide an opportunity to develop new approaches and tools that can assist So users search and (re)use architectural knowledge shared in online developer communities. To this end, in this study, we investigated architecture related questions and their associated architecture solutions in SO. Specifically, we used qualitative analysis approach to analyze a statistically representative random sample of 968 ARPs from 10,423 ARPs manually identified. We intended to identify both the categories and characteristics of architecture related questions and their solutions. We also explored the design contexts in which those questions were raised. Finally, we studied SO users’ discussions on the usefulness of the architecture solutions. We summarize our main results and findings as follows:

\begin{itemize}
\item SO users ask a broad spectrum of architecture related questions ranging from \textit{architecture tool} to \textit{architecture configuration}, \textit{architecture implementation} to \textit{architecture deployment}. In addition, SO users mostly discuss \textit{solution for architecture configuration} (39\%), followed by \textit{solution for architecture implementation} (18\%), \textit{explanation of architecture} (16\%), and \textit{architecture tactic} (11\%). We observed that ARPs (questions and answers) cover almost all architecting activities.

\item SO users ask the most (27\%, 261 out of 968) ARP questions about \textit{architecture configuration}.

\item Most of the SO users (71\%, 687 out of 968 ARP questions) considered design contexts when asking architecture related questions.

\item Architecture related questions that provide \textit{clear description together with architectural diagrams} increase their likelihood of getting more than one answer, while \textit{poorly structured architecture questions} tend to only get one answer.

\item \textit{Architecture solution for configuration} from our proposed taxonomy is the most provided type of architecture solutions that are considered useful in SO. 

\item SO users mainly consider architecture solutions that are \textit{complete and comprehensive} and have \textit{concise explanation with architectural diagrams} to be helpful.
\end{itemize}

Our results and findings can help researchers and practitioners by knowing what types of architectural knowledge, such as categories of architecture related questions and solutions, are provided in SO, and what are the characteristics of good architecture related questions and useful architecture solutions. Also, our results can motivate researchers and practitioners to consider SO as a valuable source of architectural knowledge (e.g., architecture patterns and tactics) and develop novel approaches and tools for mining useful architecture knowledge from SO to support architecting activities and development. 

In the next step: (1) We plan to conduct a comparative study of architecture solutions provided at SO and other platforms (e.g., developer mailing lists and issue tracking systems), which may help reveal insights into the current focus of architecture solutions utilization, and their advantages and deficiencies. (2) We aim for validating and extending the proposed taxonomy of useful architecture solutions provided at SO (see Figure \ref{TaxonomyOfArchSolution}) using an industrial survey from the practitioners’ perspective. (3) We also plan to design and employ (semi-)automatic approaches to extract and summarize architectural information, and establish the architecture issue-solution pairs from the retrieved architectural information, for example, benefits and drawbacks of certain architecture solutions (e.g., patterns and tactics) for task-specific architecture problems from multiple sources of architectural information (e.g., Q\&A sites, GitHub, issue tracking systems, technical blogs), which can facilitate the decision-making of architects by utilizing architectural knowledge from peers and communities. 

\section*{Acknowledgements}
This work is partially sponsored by the National Natural Science Foundation of China (NSFC) under Grant No. 62172311. The authors would also like to acknowledge the financial support from the China Scholarship Council.

\bibliography{mybibfile}

\begin{thebibliography}{10}
\expandafter\ifx\csname url\endcsname\relax
  \def\url#1{\texttt{#1}}\fi
\expandafter\ifx\csname urlprefix\endcsname\relax\def\urlprefix{URL }\fi
\expandafter\ifx\csname href\endcsname\relax
  \def\href#1#2{#2} \def\path#1{#1}\fi

\bibitem{sadowski2015developers}
C.~Sadowski, K.~T. Stolee, S.~Elbaum, How developers search for code: a case
  study, in: Proceedings of the 10th Joint Meeting of the European Software
  Engineering Conference and the ACM SIGSOFT Symposium on the Foundations of
  Software Engineering (ESEC/FSE), Bergamo, Italy, 2015, pp. 191--201.

\bibitem{zagalsky2018r}
A.~Zagalsky, D.~M. German, M.-A. Storey, C.~G. Teshima, G.~Poo-Caama{\~n}o, How
  the r community creates and curates knowledge: an extended study of stack
  overflow and mailing lists, Empirical Software Engineering 23~(2) (2018)
  953--986.

\bibitem{treude2011programmers}
C.~Treude, O.~Barzilay, M.-A. Storey, How do programmers ask and answer
  questions on the web? ({NIER Track}), in: Proceedings of the 33rd
  International Conference on Software Engineering (ICSE), Honolulu, Hawaii,
  USA, 2011, pp. 804--807.

\bibitem{musenga2022}
M.~J. de~Dieu, P.~Liang, M.~Shahin, How do developers search for architectural
  information? an industrial survey, in: Proceeding of the 19th International
  Conference on Software Architecture (ICSA), Honolulu, Hawaii, USA, 2022, pp.
  58--68.

\bibitem{bi2021mat}
T.~Bi, P.~Liang, A.~Tang, X.~Xia, Mining architecture tactics and quality
  attributes knowledge in {Stack Overflow}, Journal of Systems and Software 180
  (2021) 111005.

\bibitem{soliman2016architectural}
M.~Soliman, M.~Galster, A.~R. Salama, M.~Riebisch, Architectural knowledge for
  technology decisions in developer communities: An exploratory study with
  {StackOverflow}, in: Proceedings of the 13th Working IEEE/IFIP Conference on
  Software Architecture (WICSA), Venice, Italy, 2016, pp. 128--133.

\bibitem{diamantopoulos2015employing}
T.~Diamantopoulos, A.~Symeonidis, Employing source code information to improve
  question-answering in {Stack Overflow}, in: Proceedings of the 12th IEEE/ACM
  Working Conference on Mining Software Repositories (MSR), Florence, Italy,
  2015, pp. 454--457.

\bibitem{zhang2018code}
T.~Zhang, G.~Upadhyaya, A.~Reinhardt, H.~Rajan, M.~Kimm, Are code examples on
  an online {Q\&A} forum reliable?: A study of {API} misuse on {Stack
  Overflow}, in: Proceedings of the 40th IEEE/ACM International Conference on
  Software Engineering (ICSE), Gothenburg, Sweden, 2018, pp. 886--896.

\bibitem{liu2020mining}
D.~Liu, Z.-L. Ren, Z.-T. Long, G.-J. Gao, H.~Jiang, Mining design pattern use
  scenarios and related design pattern pairs: A case study on online posts,
  Journal of Computer Science and Technology 35~(5) (2020) 963--978.

\bibitem{soliman2018improving}
M.~Soliman, A.~R. Salama, M.~Galster, O.~Zimmermann, M.~Riebisch, Improving the
  search for architecture knowledge in online developer communities, in:
  Proceedings of the 15th IEEE International Conference on Software
  Architecture (ICSA), Seattle, WA, USA, 2018, pp. 186--195.

\bibitem{soliman2021exploring}
M.~Soliman, M.~Wiese, Y.~Li, M.~Riebisch, P.~Avgeriou, Exploring web search
  engines to find architectural knowledge, in: Proceedings of the 18th IEEE
  International Conference on Software Architecture (ICSA), Stuttgart, Germany,
  2021, pp. 162--172.

\bibitem{cervantes2016designing}
H.~Cervantes, R.~Kazman, Designing software architectures: a practical
  approach, Addison-Wesley Professional, 2016.

\bibitem{malavolta2021mining}
I.~Malavolta, K.~Chinnappan, S.~Swanborn, G.~A. Lewis, P.~Lago, Mining the ros
  ecosystem for green architectural tactics in robotics and an empirical
  evaluation, in: Proceedings of the 18th IEEE/ACM International Conference on
  Mining Software Repositories (MSR), Madrid, Spain, 2021, pp. 300--311.

\bibitem{tian2019developers}
F.~Tian, P.~Liang, M.~A. Babar, How developers discuss architecture smells?
  {An} exploratory study on {Stack Overflow}, in: Proceedings of the 16th IEEE
  International Conference on Software Architecture (ICSA), Hamburg, Germany,
  2019, pp. 91--100.

\bibitem{zhu2009multi}
Z.~Zhu, D.~Bernhard, I.~Gurevych, A multi-dimensional model for assessing the
  quality of answers in social {Q\&A} sites, in: Proceedings of the 14th
  International Conference on Information Quality (ICIQ), Potsdam, Germany,
  2009, pp. 264--265.

\bibitem{israel1992determining}
G.~D. Israel, Determining sample size, Fact Sheet PEOD-6, Florida, USA (1992).

\bibitem{stol2016grounded}
K.~S.~P. Ralph, F.~Brian, Grounded theory in software engineering research: A
  critical review and guidelines, in: Proceedings of the 38th IEEE/ACM
  International Conference on Software Engineering (ICSE), Austin, TX, USA,
  2016, pp. 120--131.

\bibitem{HofmeisterGenModel2007}
H.~Christine, P.~Kruchten, R.~L. Nord, H.~Obbink, A.~Ran, P.~America, A general
  model of software architecture design derived from five industrial
  approaches, Journal of Systems and Software 80~(1) (2007) 106--126.

\bibitem{tang2010comparative}
A.~Tang, P.~Avgeriou, A.~Jansen, R.~Capilla, M.~A. Babar, A comparative study
  of architecture knowledge management tools, Journal of Systems and Software
  83~(3) (2010) 352--370.

\bibitem{hofmeister2007general}
C.~Hofmeister, P.~Kruchten, R.~L. Nord, H.~Obbink, A.~Ran, P.~America, A
  general model of software architecture design derived from five industrial
  approaches, Journal of Systems and Software 80~(1) (2007) 106--126.

\bibitem{li2013application}
Z.~Li, P.~Liang, P.~Avgeriou, Application of knowledge-based approaches in
  software architecture: A systematic mapping study, Information and Software
  Technology 55~(5) (2013) 777--794.

\bibitem{SA2012}
L.~Bass, P.~Clements, R.~Kazman, Software Architecture in Practice, 3rd
  Edition, Addson-Wesley Professional, 2012.

\bibitem{10yearsSAKM}
C.~Rafael, A.~Jansen, A.~Tang, P.~Avgeriou, M.~A. Babar, 10 years of software
  architecture knowledge management: Practice and future, Journal of Systems
  and Software 116 (2017) 191--205.

\bibitem{jansen2005SoftArch}
A.~Jansen, J.~Bosch, Software architecture as a set of architectural design
  decisions, in: Proceedings of the 5th IEEE/IFIP Working Conference on
  Software Architecture (WICSA), Pittsburgh, Pennsylvania, USA, 2005, pp.
  109--120.

\bibitem{Malavolta2013WhatIN}
I.~Malavolta, P.~Lago, H.~Muccini, P.~Pelliccione, A.~Tang, What industry needs
  from architectural languages: A survey, IEEE Transactions on Software
  Engineering 39 (2013) 869--891.

\bibitem{bi2021architecture}
T.~Bi, W.~Ding, P.~Liang, A.~Tang, Architecture information communication in
  two oss projects: The why, who, when, and what, Journal of Systems and
  Software 181 (2021) 111035.

\bibitem{bedjeti2017modeling}
A.~Bedjeti, P.~Lago, G.~A. Lewis, R.~D.~D. Boer, R.~Hilliard, Modeling context
  with an architecture viewpoint, in: Proceedings of the 14th IEEE
  International Conference on Software Architecture (ICSA), Gothenburg, Sweden,
  2017, pp. 117--120.

\bibitem{tang2008towards}
A.~Tang, F.-C. Kuo, M.~F. Lau, Towards independent software architecture
  review, in: Proceedings of the 2nd European Conference on Software
  Architecture (ECSA), Paphos, Cyprus, 2008, pp. 306--313.

\bibitem{harper2015exploring}
K.~E. Harper, J.~Zheng, Exploring software architecture context, in:
  Proceedings of the 12th Working IEEE/IFIP Conference on Software Architecture
  (WICSA), Montréal, Québec, Canada, 2015, pp. 123--126.

\bibitem{petersen2009context}
K.~Petersen, C.~Wohlin, Context in industrial software engineering research,
  in: Proceedings of the 3rd International Symposium on Empirical Software
  Engineering and Measurement (ESEM), Lake Buena Vista, Florida, USA, 2009, pp.
  401--404.

\bibitem{groher2015study}
I.~Groher, R.~Weinreich, A study on architectural decision-making in context,
  in: Proceedings of the 12th IEEE/IFIP Working Conference on Software
  Architecture (WICSA), Montreal, QC, Canada, 2015, pp. 11--20.

\bibitem{buschmann1996pattern}
F.~Buschmann, R.~Meunier, H.~Rohnert, P.~Sommerlad, M.~Stal, P.~Sommerlad,
  M.~Stal, Pattern-Oriented Software Architecture, Vol.~1, John Wiley \& Sons,
  1996.

\bibitem{caldiera1994goal}
V.~R. Basili, G.~Caldiera, H.~D. Rombach, The goal question metric approach,
  Encyclopedia of Software Engineering (1994) 528--532.

\bibitem{soliman2015enriching}
M.~Soliman, M.~Riebisch, U.~Zdun, Enriching architecture knowledge with
  technology design decisions, in: Proceedings of the 12th Working IEEE/IFIP
  Conference on Software Architecture, (WICSA), Montreal, QC, Canada, 2015, pp.
  135--144.

\bibitem{WritingGoodquest2018}
F.~Calefatoa, F.~Lanubileb, N.~Novielli, How to ask for technical help?
  {Evidence-based} guidelines for writing questions on {Stack Overflow},
  Information and Software Technology 94 (2018) 186–207.

\bibitem{barua2014developers}
A.~Barua, S.~W. Thomas, A.~E. Hassan, What are developers talking about? an
  analysis of topics and trends in {Stack Overflow}, Empirical Software
  Engineering 19~(3) (2014) 19--32.

\bibitem{tahir2018can}
A.~Tahir, A.~Yamashita, S.~Licorish, J.~Dietrich, S.~Counsell, Can you tell me
  if it smells? a study on how developers discuss code smells and anti-patterns
  in {Stack Overflow}, in: Proceedings of the 22nd International Conference on
  Evaluation and Assessment in Software Engineering (EASE), Montreal Quebec,
  Canada, 2018, pp. 68--78.

\bibitem{allamanis2013and}
A.~Anderson, D.~Huttenlocher, J.~Kleinberg, J.~Leskovec, Discovering value from
  community activity on focused question answering sites: A case study of
  {Stack Overflow}, in: Proceeding of the 10th Working Conference on Mining
  Software Repositories (MSR), Beijing, China, 2013, pp. 53--56.

\bibitem{dataset}
J.~de~Dieu~Musengamana, P.~Liang, M.~Shahin, A.~A. Khan, Replication package
  for the paper: Characterizing architecture related posts and their usefulness
  in {Stack Overflow}, \url{https://doi.org/10.5281/zenodo.4683744}, 2022.

\bibitem{UnderstQuestQuali2014}
L.~Ponzanelli, A.~Mocci, A.~Bacchelli, M.~Lanza, Understanding and classifying
  the quality of technical forum questions, in: Proceedings of the 14th IEEE
  International Conference on Quality Software (QSIC), Allen, TX, USA, 2014,
  pp. 343--352.

\bibitem{obsolete2019}
H.~Zhang, S.~Wang, T.~P. Chen, Y.~Zou, A.~E. Hassan, An empirical study of
  obsolete answers on {Stack Overflow}, IEEE Transactions on Software
  Engineering 47~(4) (2019) 850--862.

\bibitem{readingAnswers2019}
H.~Zhang, S.~Wang, T.-H. Chen, A.~E. Hassan, Reading answers on {Stack
  Overflow}: Not enough!, IEEE Transactions on Software Engineering 47~(11)
  (2021) 2520--2533.

\bibitem{obie2022violation}
H.~O. Obie, I.~Ilekura, H.~Du, M.~Shahin, J.~Grundy, L.~Li, J.~Whittle,
  B.~Turhan, On the violation of honesty in mobile apps: Automated detection
  and categories, in: Proceedings of the 19th Working Conference on Mining
  Software Repositories (MSR), Pittsburgh, PA, USA, 2022, pp. 321--332.

\bibitem{cohen1960coefficient}
J.~Cohen, A coefficient of agreement for nominal scales, Educational and
  psychological measurement 20~(1) (1960) 37--46.

\bibitem{campbell2013coding}
J.~L. Campbell, C.~Quincy, J.~Osserman, O.~K. Pedersen, Coding in-depth
  semistructured interviews: Problems of unitization and intercoder reliability
  and agreement, Sociological Methods \& Research 42~(3) (2013) 294--320.

\bibitem{kruchten2004ontology}
P.~Kruchten, An ontology of architectural design decisions in
  software-intensive systems, in: Proceedings of the 2nd Groningen Workshop on
  Software Variability Management (SVM), Rijksuniversiteit Groningen, 2004, pp.
  54--61.

\bibitem{foote1997big}
B.~Foote, J.~Yoder, Big ball of mud, Pattern Languages of Program Design 4
  (1997) 654--692.

\bibitem{neto2005toward}
R.~de~Freitas Bulcao~Neto, M.~da~Graca Campos~Pimentel, Toward a
  domain-independent semantic model for context-aware computing, in: Proceeding
  of the 3rd Latin American Web Congress (LA-WEB), Buenos Aires, Argentina,
  2005, pp. 10--19.

\bibitem{petrov2011need}
P.~Petrov, U.~Buy, R.~L. Nord, The need for a multilevel context-aware software
  architecture analysis and design method with enterprise and system
  architecture concerns as first class entities, in: Proceedings of the 9th
  Working IEEE/IFIP Conference on Software Architecture (WICSA), Boulder,
  Colorado, USA, 2011, pp. 147--156.

\bibitem{asaduzzaman2013answering}
M.~Asaduzzaman, A.~S. Mashiyat, C.~K. Roy, K.~A. Schneider, Answering questions
  about unanswered questions of {Stack Overflow}, in: Proceedings of the 10th
  Working Conference on Mining Software Repositories (MSR), San Francisco, CA,
  USA, 2013, pp. 97--100.

\bibitem{soliman2017developing}
M.~Soliman, M.~Galster, M.~Riebisch, Developing an ontology for architecture
  knowledge from developer communities, in: Proceedings of the 14th IEEE
  International Conference on Software Architecture (ICSA), Gothenburg, Sweden,
  2017, pp. 89--92.

\bibitem{bi2018architecture}
T.~Bi, P.~Liang, A.~Tang, Architecture patterns, quality attributes, and design
  contexts: How developers design with them?, in: Proceedings of the 25th
  Asia-Pacific Software Engineering Conference (APSEC), Nara, Japan, 2018, pp.
  49--58.

\bibitem{wang2018users}
S.~Wang, T.~P. Chen, A.~E. Hassan, How do users revise answers on technical
  {Q\&A} websites? {A} case study on {Stack Overflow}, IEEE Transactions on
  Software Engineering 46~(3) (2020) 1024--1038.

\bibitem{GoodAcode2012}
S.~M. Nasehi, J.~Sillito, F.~Maurer, C.~Burns, What makes a good code example?
  a study of programming {Q\&A} in {StackOverflow}, in: Proceedings of the 28th
  IEEE International Conference on Software Maintenance (ICSM), Trento, Italy,
  2012, pp. 25--34.

\bibitem{zhang2021comments}
H.~Zhang, S.~Wang, T.~P. Chen, A.~E. Hassan, Are comments on {Stack Overflow}
  well organized for easy retrieval by developers?, ACM Transactions on
  Software Engineering and Methodology 30~(2) (2021) Article No. 22.

\bibitem{nadi2020essential}
S.~Nadi, C.~Treude, Essential sentences for navigating {Stack Overflow}
  answers, in: Proceedings of the 27th IEEE International Conference on
  Software Analysis, Evolution and Reengineering (SANER), London, ON, Canada,
  2020, pp. 229--239.

\bibitem{thomas2013con}
T.~Haitzer, U.~Zdun, Controlled experiment on the supportive effect of
  architectural component diagrams for design understanding of novice
  architects, in: Proceedings of the 7th European Conference on Software
  Architecture (ECSA), Montpellier, France, 2013, pp. 54--71.

\bibitem{GoodAnsw2012}
Y.~Yao, H.~Tong, T.~Xie, L.~Akoglu, F.~Xu, J.~Lun, Want a good answer? ask a
  good question first! (2013), arXiv:1311.6876.

\bibitem{wijerathna2022mining}
L.~Wijerathna, A.~Aleti, T.~Bi, A.~Tang, Mining and relating design contexts
  and design patterns from {Stack Overflow}, Empirical Software Engineering
  27~(1) (2022) 1--53.

\bibitem{ref35}
C.~Wohlin, P.~Runeson, M.~H{\"o}st, M.~C. Ohlsson, B.~Regnell, A.~Wessl{\'e}n,
  Experimentation in Software Engineering, Springer, 2012.

\bibitem{chinnappan2021architectural}
K.~Chinnappan, I.~Malavolta, G.~A. Lewis, M.~Albonico, P.~Lago, Architectural
  tactics for energy-aware robotics software: A preliminary study, in:
  Proceedings of the 15th European Conference on Software Architecture (ECSA),
  Virtual Event, Sweden, 2021, pp. 164--171.

\bibitem{li2021understanding}
R.~Li, P.~Liang, M.~Soliman, P.~Avgeriou, Understanding architecture erosion:
  The practitioners’ perceptive, in: Proceeding of the 29th IEEE/ACM
  International Conference on Program Comprehension (ICPC), Madrid, Spain,
  2021, pp. 311--322.

\bibitem{zou2017towards}
J.~Zou, L.~Xu, M.~Yang, X.~Zhang, D.~Yang, Towards comprehending the
  non-functional requirements through developers’ eyes: An exploration of
  {Stack Overflow} using topic analysis, Information and Software Technology 84
  (2017) 19--32.

\bibitem{dagenais2012recovering}
B.~Dagenais, M.~P. Robillard, Recovering traceability links between an {API}
  and its learning resources, in: Proceedings of the 34th IEEE International
  Conference on Software Engineering (ICSE), Zurich, Switzerland, 2012, pp.
  47--57.

\bibitem{an2017stack}
L.~An, O.~Mlouki, F.~Khomh, G.~Antoniol, {Stack Overflow}: A code laundering
  platform, in: Proceedings of the 24th IEEE International Conference on
  Software Analysis, Evolution and Reengineering (SANER), Klagenfurt, Austria,
  2017, pp. 283--293.

\bibitem{fischer2017stack}
F.~Fischer, K.~B{\"o}ttinge, H.~Xiao, C.~Stransky, Y.~Acar, M.~Backes, S.~Fahl,
  {Stack Overflow} considered harmful? the impact of copy\&paste on android
  application security, in: Proceeding of the 38th IEEE Symposium on Security
  and Privacy (S\&P), San Jose, CA, USA, 2017, pp. 121--136.

\bibitem{ragkhitwetsagul2019toxic}
C.~Ragkhitwetsagul, J.~Krinke, M.~Paixao, G.~Bianco, R.~Oliveto, Toxic code
  snippets on stack overflow, IEEE Transactions on Software Engineering 47~(3)
  (2019) 560--581.

\bibitem{zagalsky2012example}
A.~Zagalsky, O.~Barzilay, A.~Yehudai, Example overflow: Using social media for
  code recommendation, in: Proceedings of the 3rd International Workshop on
  Recommendation Systems for Software Engineering (RSSE), Zurich, Switzerland,
  2012, pp. 38--42.

\bibitem{treude2017understanding}
C.~Treude, M.~P. Robillard, Understanding {Stack Overflow} code fragments, in:
  Proceedings of the 33rd IEEE International Conference on Software Maintenance
  and Evolution (ICSME), Shanghai, China, 2017, pp. 509--513.

\bibitem{agkhitwetsagul2018awareness}
C.~Ragkhitwetsagul, J.~Krinke, R.~Oliveto, Awareness and experience of
  developers to outdated and license-violating code on {Stack Overflow}: An
  online survey (2018), arXiv:1806.08149.

\bibitem{treude2016augmenting}
C.~Treude, M.~P. Robillard, Augmenting {API} documentation with insights from
  {Stack Overflow}, in: Proceedings of the 38th International Conference on
  Software Engineering (ICSE), Austin, Texas, USA, 2016, pp. 392--403.

\bibitem{wong2013autocomment}
E.~Wong, J.~Yang, L.~Tan, Autocomment: Mining question and answer sites for
  automatic comment generation, in: Proceedings of the 28th IEEE/ACM
  International Conference on Automated Software Engineering (ASE), Silicon
  Valley, CA, USA, 2013, pp. 562--567.

\bibitem{gao2015fixing}
Q.~Gao, H.~Zhang, J.~Wang, Y.~Xiong, L.~Zhang, H.~Mei, Fixing recurring crash
  bugs via analyzing {Q\&A} sites, in: Proceedings of the 30th International
  Conference on Automated Software Engineering (ASE), Lincoln, NE, USA, 2015,
  pp. 307--318.

\bibitem{mcdonnell2013empirical}
T.~McDonnell, B.~Ray, M.~Kim, An empirical study of {API} stability and
  adoption in the android ecosystem, in: Proceedings of the 29th IEEE
  International Conference on Software Maintenance (ICSM), Eindhoven, The
  Netherlands, 2013, pp. 70--79.

\bibitem{dalip2013exploiting}
D.~H. Dalip, M.~Cristo, P.~Calado, Exploiting user feedback to learn to rank
  answers in {Q\&A} forums: A case study with {Stack Overflow}, in: Proceedings
  of the 36th International ACM SIGIR Conference on Research and Development in
  Information Retrieval (SIGIR), Dublin, Ireland, 2013, pp. 543--552.

\bibitem{xu2017answerbot}
B.~Xu, Z.~Xing, X.~Xia, D.~Lo, Answerbot: Automated generation of answer
  summary to developers’ technical questions, in: Proceedings of the 32nd
  IEEE/ACM International Conference on Automated Software Engineering (ASE),
  Urbana, IL, USA, 2017, pp. 706--716.

\bibitem{calefato2019empirical}
F.~Calefato, F.~Lanubile, N.~Novielli, An empirical assessment of best-answer
  prediction models in technical {Q\&A} sites, Empirical Software Engineering
  24~(2) (2019) 854--901.

\end{thebibliography}
\end{sloppypar}
\end{document}